\documentclass[conference]{IEEEtran}
\usepackage{cite}
\usepackage{amsmath,amssymb,amsfonts}
\usepackage{algorithmic}
\usepackage{graphicx}
\usepackage{textcomp}
\usepackage{xcolor}
\usepackage{listings}
\usepackage{comment}
\usepackage{amsthm}

\usepackage{float}
\usepackage[]{algorithm2e}

\usepackage{pifont}

\usepackage{tikz}
\usetikzlibrary{trees,automata,positioning,shapes,shadows,arrows,calc}

\usepackage{multirow}
\newif\ifclean
\cleantrue
\newif\ifnew
\newfalse
\cleanfalse

\newtheorem{definition}{Definition}

\makeatletter

\setbox0\hbox{$\xdef\scriptratio{\strip@pt\dimexpr
    \numexpr(\sf@size*65536)/\f@size sp}$}

\newcommand{\scriptveryshortarrow}[1][3pt]{{%
    \vcenter{\hbox{\rule[\scriptratio\dimexpr-.2pt\relax]
               {\scriptratio\dimexpr#1\relax}{\scriptratio\dimexpr.4pt\relax}}}%
   \mkern-4mu\hbox{\let\f@size\sf@size\usefont{U}{lasy}{m}{n}\symbol{41}}}}

\makeatother

\newcommand{\tuple}[1]{\langle #1\rangle}

\newcommand{\senc}{\mathsf{senc}}
\newcommand{\sign}{\mathsf{sign}}
\newcommand{\aenc}{\mathsf{aenc}}

\newcommand{\m}[1]{\textit{#1}} % cryptographic messages
\newcommand{\cst}[1]{\texttt{'#1'}} % constants 
\newcommand{\ltk}{\m{ltk}}
\newcommand{\ltkA}{\ltk_\m{1}}
\newcommand{\ltkB}{\ltk_\m{2}}
\newcommand{\ltkC}{\ltk_\m{3}}
\newcommand{\ltkN}{\ltk_\m{n}}
\newcommand{\ltkNp}{\ltk_\m{n+1}}

\newcommand{\ltkCA}{\ltk_\m{CA}}

\newcommand{\jrek}{\m{jrek}}

\newcommand{\pkjrek}{\mathsf{pk}(\jrek)}

\newcommand{\ejoin}{\m{eJoin}}

\newcommand{\eLeave}{\m{eLeave}}
\newcommand{\eKUR}{\m{eKUR}}
\newcommand{\eKupdate}{\m{eKupdate}}

\newcommand{\pgkUpdate}{\m{pgkUpdate}}

\newcommand{\ppkA}{\m{ppk}_1}
\newcommand{\ppkB}{\m{ppk}_2}
\newcommand{\pgk}{\m{pgk}}
\newcommand{\ppkN}{\m{ppk}_n}

\newcommand{\rekRecipInfo}{\m{rekRecipInfo}}
\newcommand{\symmRecipInfo}{\m{symmRecipInfo}}
\newcommand{\ciphertext}{\m{ciphertext}}
\newcommand{\certA}{\m{Cert}_\m{1}}
\newcommand{\certB}{\m{Cert}_\m{2}}
\newcommand{\certC}{\m{Cert}_\m{3}}
\newcommand{\certN}{\m{Cert}_\m{n}}
\newcommand{\certNp}{\m{Cert}_\m{n+1}}

\newcommand{\APP}{\m{APP}}
\newcommand{\GN}{\m{GN}}
\newcommand{\LEAVEMSG}{\m{leaveMsg}}

\newcommand{\idC}{\m{$3$}}
\newcommand{\idN}{\m{$n$}}

\newcommand{\platoonID}{\m{P}}
\newcommand{\pos}{\m{position}}

\newcommand{\CAM}{\cst{CAM}}
\newcommand{\LEAVE}{\cst{Leave}}

\newcommand{\reason}{\cst{Reason}}

\newcommand{\KUR}{\cst{KUR}}
\newcommand{\KUpdate}{\cst{KeyUpdate}}

\newcommand{\JR}{\cst{JoinRequest}}
\newcommand{\JRE}{\cst{JoinResponse}}

\newcommand{\IEEESTD}{\normalfont{IEEE 1609.2}\xspace}
\newcommand{\ETSISEC}{\normalfont{ETSI 103 097}\xspace}

\renewcommand{\quote}[3]{\begin{formal}\textcolor{black}{\bf [\texttt{#1}]} #3\end{formal}}  % Arg 1: TS/TR document, Arg 2: Section of the doc, Arg3: Quote

\newcommand{\quoteIEEE}[2]{\quote{IEEE\,1609.2}{#1}{#2}}  % Arg 1: Section of the TS 33.501 doc., Arg2: Quote

\newcommand{\messLabel}[1]{\small{#1}}

\newcommand{\tamarin}{\textsc{Tamarin}\xspace}
\newcommand{\proverif}{\textsc{ProVerif}\xspace}

\newcommand{\scyther}{\textsc{Scyther}\xspace}
\newcommand{\rewire}{\textsc{Rewire}\xspace}

\newcommand{\factStyle}[1]{\textsf{#1}}

\usepackage{array}
\newcolumntype{P}[1]{>{\centering\arraybackslash}p{#1}}

\usepackage{listings}

\usepackage{latexsym,amsthm}
\usepackage{xspace}
\usepackage{enumitem}           % mainly for (i),(ii) in enumerate env.

\usepackage[strict]{changepage} % for adjustwidth environment
\usepackage{framed} % for formal definitions

\usepackage{msc}
\usepackage{anb}
\usepackage{booktabs}           % better tabular
\usepackage[most]{tcolorbox}    % for tcolorbox

\usepackage{todonotes}
\usepackage{pdfcomment}

\usepackage{cleveref}

\definecolor{formalshade}{rgb}{0.95,0.95,1}
\definecolor{darkblue}{rgb}{0.0, 0.0, 0.55}
\definecolor{seagreen}{rgb}{0.18, 0.55, 0.34}
\definecolor{applegreen}{rgb}{0.55, 0.71, 0.0}
\definecolor{asparagus}{rgb}{0.53, 0.66, 0.42}
\definecolor{black}{rgb}{0.0, 0.0, 0.0}
\definecolor{grannysmithapple}{rgb}{0.66, 0.89, 0.63}
\definecolor{mediumseagreen}{rgb}{0.24, 0.7, 0.44}
\definecolor{palegold}{rgb}{0.9, 0.75, 0.54}
\definecolor{palegoldenrod}{rgb}{0.93, 0.91, 0.67}
\newenvironment{formal}{%
\MakeFramed{\advance\hsize-\width\FrameRestore}%
\noindent\hspace{-4.55pt}% disable indenting first paragraph
\begin{adjustwidth}{}{7pt}%
\vspace{2pt}\vspace{2pt}%
}
{%
\vspace{2pt}\end{adjustwidth}\endMakeFramed%
}

\begin{document}

\title{Exploiting Partial Order of Keys to Verify Security of a Vehicular Group Protocol}

\author{\IEEEauthorblockN{Felipe Boeira and Mikael Asplund} \IEEEauthorblockA{Dept. of Computer and Information Science\\Link\"{o}ping University, Sweden}}

\maketitle

%fix notation # and $
%add text to ASN.1 for data checks
%fix the flow of protocol security verification
% DESCRIBE PROTOCOL AS key establishment and distribution

\begin{abstract}
Vehicular networks will enable a range of novel applications to enhance road traffic efficiency, safety, and reduce fuel consumption. As for other cyber-physical systems, security is essential to the deployment of these applications and standardisation efforts are ongoing. In this paper, we perform a systematic security evaluation of a vehicular platooning protocol through a thorough analysis of the protocol and security standards. We tackle the complexity of the resulting model with a proof strategy based on a relation on keys. The key relation forms a partial order, which encapsulates both secrecy and authenticity dependencies. We show that our order-aware approach makes the verification feasible and proves authenticity properties along with secrecy of all keys used throughout the protocol.
\end{abstract}

%In recent years, standardization entities have released security technical specifications and standards to support the development of secure vehicular applications. However, there has been insufficient research in formally analyzing security aspects of employing such standards in the development of vehicular protocols. In this paper, we formally analyze the utilization of IEEE and ETSI standards in the context of Ensemble - an European initiative to realize multi-brand truck platooning.

%CPS/vehicles
%standardisation/security is important
%systematic sec evaluation of a platooning protocol
%tackle complexity by analyzing key relations
%proof strategy
%results    

\section{Introduction}

Security of cyber-physical systems are increasingly becoming a societal concern, as both the attack surfaces and the potential consequences of attacks have increased considerably in recent years. In the automotive domain, connected and automated vehicles (CAV) promise considerable improvements in safety and efficiency, but also require very strict security processes to be trustworthy. Today, the most advanced and complex collaborative vehicular application comes in the form of vehicular platooning where a group of vehicles are jointly controlled by a leader. In this paper we consider the security of platooning as a starting point to investigate automated security proofs, partially ordered key structures, and the process of transforming informal and semi-formal standardisation text to formal models.

Our work was inspired by Basin et al.~\cite{10.1145/3243734.3243846} who in their security analysis of the 5G AKA protocol make a compelling argument for the need of formal models and security specifications to complement protocol standards. In particular, the authors point to under-specification of security properties and assumptions that can in some cases lead to vulnerabilities. Our analysis is focused on a cyber-physical protocol that is currently in the pre-standardisation phase and described in the European Ensemble project, and which also builds on the existing ETSI ITS-G5 and IEEE vehicular networking standards (including security). Together, these form an interesting study object since (i) they will have a real and significant impact on the way future commercial vehicles are operated and controlled, (ii) they represent a typical standardisation product composed of multiple cross-references documents (in our case 8 documents and 617 pages), and (iii) the protocol and the associated security specification describe a complex system with dynamically joining and leaving nodes and a non-trivial key structure.

We perform a structured and formal security analysis of the Ensemble platooning protocol. To perform this analysis several interesting challenges must be overcome. One such challenge is the transformation from informal and semi-formal standard documents - where descriptions are often spread out over multiple documents, contain optional parts, and sometimes overrides previous statements - to a formal model. We show how ASN.1 specifications significantly improve this process and discuss potential benefits of a fully automatic transformation process. Moreover, our analysis also identifies the lack of a corresponding language to capture the behaviour of security checks since the specifications currently included only capture the \emph{structure} of security information, not the \emph{mechanisms} that make use of that information. This gives rise to potential weaknesses depending on the interpretation of the standard.

The second major challenge is of course the model complexity which causes a state space explosion. In our case, even a simple fact such as the secrecy of a long-term key could not be proven on a large computing cluster without manual intervention. Previous work have shown how this process can be aided with so called helper lemmas and oracles. The difficulty with such mechanisms is that they are inherently specific to the problem at hand, and hard to generalise. In this work we explore how the structure of the model can be used to guide the proof strategy. In particular, by considering the ordered structure of the cryptographic keys in the model we make the problem tractable, thereby allowing a more generic proof guidance mechanism. 

Secure protocols are often created so that multiple secret keys form dependency chains where the secrecy of one key is dependent on the secrecy of another. This naturally allows formulating provable properties in a way that together satisfy the overall security specification (e.g., see ~\cite{paulson01}). However, we are not aware of any previous work that formalises the two forms of dependencies (authenticity and secrecy dependencies) into a partial order on keys and translate it into a security verification proof strategy. 

%  With our dependency relation, these classes of keys form a partial order and we demonstrate that by structuring the proof strategy as a linear extension of this partial order, we can prove secrecy of all keys and certain authenticity properties. 

We formulate relevant security properties of vehicular communication protocols and instantiate our model and proof guidance\footnote{The full model will be made available in a public repository, and is provided as supplementary material for the reviewers.} in the \tamarin verifier tool~\cite{10.1007/978-3-642-39799-8_48}. In our case study there are 10 classes of keys with up to a dependency depth of six, and potentially infinitely many instances of keys on a protocol run. 
%We discuss an interesting aspect of key secrecy that concerns shared keys in dynamic groups, and which highlights a fundamental problem in collaborative protocols with open participation.
Our improved proof strategy allows us to identify the security properties that are met by current protocols and under what circumstances. Overall, our assessment is that security standards for vehicular networks can provide strong security properties, but that ambiguities and implicit assumptions in the standards potentially give room for implementations with lacking security checks.

To summarise, the contributions of this paper are as follows.

\begin{itemize}
    \item Formulation of a joint secrecy and authenticity relation on a set of keys that potentially forms a partial order, together with an automated proof strategy, a key hierarchy extractor for \tamarin protocol models that exhibit such partial orders of keys. 
    \item An assessment of the state of security for current vehicular networking protocols and a formal and structured security analysis of vehicular platooning.
    \item A structured approach for interpreting security standards in order to create models that allow formal reasoning, identification of what is lacking in current standards, and recommendations for future vehicular security standardisation activities.

\end{itemize}

The remainder of this paper is organised as follows: Section~\ref{sec:background} introduces the area of vehicular networking protocols at large, the Ensemble platooning protocol in particular and formal verification of security protocols. Section~\ref{sec:protocol_model} introduces our model of the platooning protocol and the security properties that we verify. Section~\ref{sec:proof_strategy} introduces the partial order of keys, and explains how this ordering relation is used to create the proof guidance. The outcome of verifying security protocols under different settings and assumptions is presented in Section~\ref{sec:results}. Finally, Section~\ref{sec:related_work} describes related work and Section~\ref{sec:conclusion} concludes the paper and outlines future work.

\section{Background}
\label{sec:background}

%We need to explain why we do this, and point to something that interests the reader
%also, we need to zoom in from a larger perspective

In this section we introduce the standards for vehicular communication and security that are employed in Ensemble. We describe the overall design of the protocol, and provide an overview of how the protocol security properties are verified using the \tamarin security verification tool.

% add motivation to symbolic model

\subsection{Vehicular network and security standards}
\label{subsec:standards}
% 0.75 page

In recent years, organisations such as the Institute of Electrical and Electronics Engineers (IEEE) and European Telecommunications Standards Institute (ETSI) have been actively working towards standardisation of vehicular network protocols and applications. The Ensemble protocol~\cite{ensemble-d8, ensemble-d9} is built on top of these existing standards and makes use of their services, so one must understand how these fit together to understand the security implications for vehicular platooning.

%vehicular standardization and a number of their deliverables are used as building blocks in the design of Ensemble.

\begin{figure}[ht]
    \centering
\begin{tikzpicture}[align=center,>=latex',font=\sffamily]
\tikzstyle{Dotted} = [draw=black,dashed,thick,rectangle,minimum width=80mm,minimum height=33mm]
\tikzstyle{Normal} = [draw=black,thick,rectangle,minimum width=68mm,minimum height=7mm]
\node[label=above:Access Layer,label={[label distance=5.5mm]},Dotted,minimum height=13mm](Controller){};
\node[above right = 3mm and 6mm of Controller.south west, Normal,minimum width=20mm](ITSG5){EN 302 663 (ITS-G5)};
\node[right = 3mm and 6mm of ITSG5.east,Normal,minimum width=20mm](V2X){3GPP V2X (5G)};

%\node[above right = 9mm and 0mm of ITSG5.west,Normal](LL){Link Layer};

\node[above = 6mm of Controller,label=above:Networking and Transport Layer,label={[label distance=5.5mm]},Dotted,minimum height=22mm](Host){};
\node[above left = 3mm and 6mm of Host.south east,anchor=south east,Normal](L2CAP){Geonetworking (GN)};

\node[above = 9mm of L2CAP.north west,anchor=north west,Normal](GAP){Basic Transport Protocol (BTP)};

\node[above =6mm of Host,label=above:Facilities Layer,label={[label distance=5.5mm]},Dotted,minimum height=13mm](Facilities){};

\node[above right = 3mm and 6mm of Facilities.south west,label={[label distance=5.5mm]},Normal,minimum width=20mm](DENM){DENM};

\node[right = 3mm  of DENM.east,label={[label distance=5.5mm]},Normal,minimum width=20mm](Application){[...]};

\node[above left = 3mm and 6mm of Facilities.south east,label={[label distance=5.5mm]},Normal,minimum width=20mm](Application){ENSEMBLE};

\node[above left = 2mm and 1mm of Controller.north west, Normal,minimum height=46mm, minimum width=7mm](IEEE){\rotatebox{90}{IEEE 1609.2 / ETSI 103 097}};

\end{tikzpicture}

\caption{ENSEMBLE protocol stack.}
    \label{fig:protocol_stack}
\end{figure}
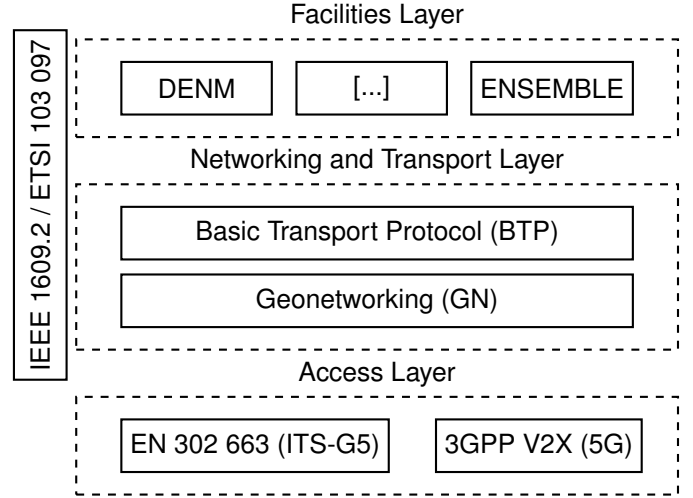

Fig.~\ref{fig:protocol_stack} represents the protocol stack and the associated standards for each layer. In Europe, the physical layer and the data link layer are grouped together into the \emph{Access layer} and are part of ITS-G5. More recently, the 5G standard has emerged as a potential replacement at the access layer. Ensemble is designed to work on ITS-G5 together with the Geonetworking (GN) protocol and the Basic Transport Protocol (BTP) which compose the network and transport layer. A set of security profiles for vehicular applications is defined by ETSI based on the primitives and message types defined by IEEE, and are applied to GN and Ensemble messages. We proceed to describe each of these standards briefly (starting from the lowest layer).

\par \textbf{ITS-G5 (ETSI EN 302 663).}
At the bottom of the vehicular networking stack we find the access layer, which is of limited interest from a security standpoint, but still relevant to provide context. In Europe and US, the dominating access layer standard for vehicular networks has been based on a flavour of IEEE 802.11 (i.e., essentially wifi) operating at 5.9GHz.

\par \textbf{3GPP V2X (5G Release 15+).} 
%Recently, there has been a shift towards using extensions of cellular standards under the 5G umbrella since it is expected that future connected vehicles will support this technology. 
There are ongoing efforts to integrate 5G as an alternative access layer in the vehicular networking stack~\cite{3GPP-r15}. Provided that GN is compatible with V2X as intended~\cite{etsi-ltev2x}, we argue that our analysis will still be relevant as the cryptographic operations occur from the networking and transport layer upwards.
    
\par \textbf{GeoNetworking (ETSI EN 302 636-4-1).} The GN protocol~\cite{etsi-gn} provides packet routing in vehicular networks with the use of geographical locations for packet transport. In Ensemble only the single hop broadcast (SHB) mode is used.

\par \textbf{BTP (ETSI EN 302 626-5-1)}. BTP~\cite{etsi-btp} defines two header variants: BTP-A for interactive packet transport and BTP-B for non-interactive. %Both headers have only two fields. While the destination port is carried by both variants, BTP-A includes a source port and BTP-B includes a destination port info, instead. 
Given that BTP does not carry relevant information for our analysis and is included in higher layers' cryptographic operations, we have decided to omit it in our model.

\par \textbf{Security Services for Applications and Management Messages (IEEE 1609.2).}
The primitives for providing security capabilities to vehicular messages are primarily defined by IEEE 1609.2~\cite{ieee-1609.2} and its two amendments~\cite{ieee-1609.2a, ieee-1609.2b}. The standard defines several data structures for encapsulating data based on the type and origin of the key that is used to encrypt it, and a \texttt{SignedData} structure for storing signature information. The 1609.2 standard uses the concept of recipients to transfer data encryption keys to other nodes, and a sequence of recipients may be included in a message. For instance, a recipient of type $\texttt{pskRecipInfo}$ is used whenever the data encryption key is pre-shared between the participants. Alternatively, a node might use an ephemeral data encryption key and a key encryption key to protect the ephemeral. In particular, two recipient types that employ ephemeral keys are used in Ensemble:

\begin{itemize}
    \item \texttt{symmRecipInfo} specifies that the ephemeral data encryption key was encrypted using a symmetric key.
    \item \texttt{rekRecipInfo} specifies that the ephemeral data encryption key was encrypted using a public response encryption key that was not obtained from a \texttt{SignedData} structure. 
\end{itemize}

\par \textbf{Security header and certificate formats (ETSI TS 103 097).}
While IEEE 1609.2 defines low-level primitives and data structures to build vehicular messages, the ETSI TS 103 097~\cite{etsi-sec} specifies message profiles based on the definitions from 1609.2. For instance, ETSI specifies that Decentralized Environmental Notification Messages (DENMs) shall include the certificate as the signer information instead of its digest only. The Ensemble project extends the definitions contained in this standard to reflect the requirements of the messages exchanged in the protocol.

%Although Ensemble has originally been specified under ITS-G5, it is possible that the V2X specification in the latest releases of 3GPP. 

\subsection{Platooning protocol}
% 0.75 page

Ensemble works as a group formation protocol with key establishment and distribution. It consists of four operational modes: \texttt{idle}, \texttt{join}, \texttt{platoon}, and \texttt{leave}. In \texttt{idle} mode, vehicles announce their interest to form a platoon through the flag \texttt{isJoinable}, which is included in a Cooperative Awareness Message (CAM). Neighbouring vehicles may send a request to join the platoon and, if accepted, will receive a join response. These two messages compose the \texttt{join} mode, which enables the joiner to start receiving and sending control messages in the \texttt{platoon} mode. The latest vehicle to join the platoon may flag \texttt{isJoinable} to allow other neighbours to join, up to a certain length of the platoon. Finally, the \texttt{leave} mode is activated whenever a vehicle wishes to depart from the platoon or if no control messages have been received for a predetermined period of time. More details on the protocol are provided in Section~\ref{sec:protocol_model} where we describe how it is formally modelled. 

\subsection{Protocol security verification}
\label{sec:protocol-sec-verif}
% 0.5 page
To formally prove security properties of a protocol such as Ensemble, there are at least three things to consider, how to model the protocol, how to model the attacker, and what security properties to verify.

%... \todo{not sure how to summarise this, need to discuss}
%Protocols must be mathematically represented in order to obtain security proofs, and 
There are two protocol models generally considered for creating cryptographic protocol representations: computational and symbolic~\cite{10.1007/978-3-642-28641-4_2}. In the computational model~\cite{10.1137/0217017, GOLDWASSER1984270}, terms are represented as bitstrings, cryptographic primitives are functions on these bitstrings, and the adversary is any probabilistic Turing machine.
In the symbolic model (which we consider in this work) bitstrings are abstracted to algebraic terms, and cryptographic primitives are represented by an equational theory. Messages are terms of these equations, for instance, consider the symmetric encryption/decryption equation where the term $m$ is encrypted/decrypted using the key $k$ in Equation~\ref{eq:senc}.

\begin{equation}
\label{eq:senc}
    sdec(senc(m, k), k) = m
\end{equation}

The symbolic model allows the reasoning to be automated, although complex protocols usually require the solver to be guided with some proof strategy as we will discuss in the later sections of this work. Given proper heuristics, \tamarin has been shown to work with protocols that exhibit complex state machines that may include loops and agent memory~\cite{10.1145/3243734.3243846, 7546518, 10.1007/978-3-662-46666-7_12, 10.1007/978-3-319-66402-6_23, cremers2020formal}. For these reasons, the symbolic model and \tamarin tool are well-suited to our work.

To represent an attacker that acts throughout the execution of the protocol, a threat model defines capabilities on computation and on observing and manipulating the network communication. A Dolev-Yao~\cite{1056650} threat model assumes a powerful attacker who is able to tamper with public communication channels, knows public constants, has unbounded computational and communication resources, and is able to employ cryptographic primitives as long as the required terms are known.

We now provide a brief overview of modelling in \tamarin. Protocols and adversary actions are modelled as multiset rewriting rules and security properties are defined through (temporal) first-order logic. Rules are composed by \textbf{p}remise, \textbf{a}ction, and \textbf{c}onclusion facts as follows: 
\begin{center}
    $[p_1, ...,p_i]\  \textnormal{--}\![a_1, ..., a_j]\!\!\!\rightarrow [c_1, ..., c_k]$
\end{center}
The solver maintains a multiset of facts that can be consumed as premises to activate the execution of a rule (there are also persistent facts that can be consumed an arbitrary number of times and are defined with a starting '!'). 
%Facts may be consumed an arbitrary number of times if annotated with an exclamation mark, and are denoted persistent facts. 
The special facts $\factStyle{In()}$ and $\factStyle{Out()}$ are used to model receiving and sending messages over the network (which can be intercepted by the attacker), and $\factStyle{Fr()}$ to generate fresh terms. In addition, the action fact $\factStyle{KU()}$ logs terms that are known by the attacker. The execution of rules creates a trace of action facts,
%that are associated with time points depending on the execution order of the rules, 
and the security properties are formulas that reason about possible traces of the protocol. For instance, consider an action fact $\factStyle{Secret}(x)$ that marks the term $x$ as secret whenever the corresponding rule is executed. The following formula formalises the property that an adversary cannot know a secret term $x$  at any time point $j$ (an action fact $a$ that occurs at time point $j$ is denoted as $a@j$).

\begin{center}
    $\forall x\ i.\ \factStyle{Secret}(x)@i \Rightarrow \lnot(\exists\ j.\ \factStyle{KU}(x)@j)$
\end{center}

To construct traces for which formulas will be checked, \tamarin uses a backwards constraint solving approach that checks all possible sources for a given constraint. For a more detailed discussion and presentation of \tamarin we refer the reader to~\cite{20.500.11850/66840, 20.500.11850/72713, 8429317, 10.1007/978-3-030-59013-0_1, dreier:hal-02358878}.

% describe the following without subsubsections
%\subsubsection{Protocol modeling}

%\subsubsection{Property specification}

%\subsubsection{Tamarin}

% 2 pages

\section{Protocol model}
\label{sec:protocol_model}

This section describes our approach to analysing the standards and interpreting semi-formal descriptions to create a formal protocol model. We describe how we leverage ASN.1 (Abstract Syntax Notation One) specifications to support this process and present the resulting models (we model the platoon formation statically and dynamically), the verification goals, and assumptions we have considered.

\subsection{Protocol messages interpretation}
%talk about ASN.1 message interpretation translation to tamarin
% 0.5 page

% talk about the problem
% get modules
% get packets
% describe structured approach 
% explicitly explain that we do not automatically convert ASN.1 to tamarin

Some of the complexities to model protocols lie in collecting information that is scattered across different documents and connecting information that is defined sparsely, as well as interpreting possibly ambiguous specifications with regards to, for example, whether to include certain optional fields in a message. In our analysis, basic data structures and message types are defined by \IEEESTD, which is extended by two amendments, and \ETSISEC defines profiles based on these definitions. Finally, the protocol specification itself uses and extends the profiles in distinct ways (Ensemble).

%Consider the case where several basic data structures with mandatory and optional fields are defined in a standard. Then, basic message types may be defined based on such basic data structures and extended to create more specific message types, which may require certain optional fields differently. In addition, amendments to the standard may change definitions (such as the case of \IEEESTD), which may modify basic data structures and impact items derived from them. Furthermore, other standardisation entities might extend those message profiles (in our case \ETSISEC defines profiles based on prior definitions from \IEEESTD), and finally, the protocol specification itself might use the profiles in distinct ways (Ensemble).

%To support the modelling process and alleviate the mentioned issues it is useful to have a complete example of what the final message types contain, as specified by the protocol. We have used ASN.1 specifications that are included in several of the standards that we analyse to guide our modelling of the protocol messages.

%ASN.1 is an abbreviation for Abstract Syntax Notation One and is an ITU/ISO standard notation widely used to define data structures in telecommunications and computer networks. It defines how data is to be encoded/decoded regardless of the programming language or physical medium by which data is transmitted.

%An overview of the the protocol modelling process is shown in Fig.~\ref{fig:asn1}. We show that, 
We have used ASN.1 specifications that are included in several of the standards that we analyse to guide our modelling of the protocol messages. First, we collect the required modules from distinct standards (data structures and message types can be defined and imported as modules): ITS Container, CAM, IEEE 1609.2, ETSI 103 097, and Ensemble. Then, we employ an ASN.1 compiler to generate sample packets of the data structures we are interested in modelling. With the final sample packet, we can refer back to the standards so that the expected behaviour of the agents towards the data included those packets can be modelled. We refine and choose what to model in a message type if a given data structure transmits or modifies cryptographic material, and whether it affects the way the agents handle the messages (e.g., the presence of a node identity in a message can be matched with expected senders by a receiver).

In this work, the transition from ASN.1 packet descriptions to a formal model was performed manually. Given that large portions of the \tamarin model are based on the content of messages as they are exchanged between different nodes, much of this process should be possible to automate. If properly implemented, this would significantly simplify the process of model validation.

%\begin{figure}[h]
%    \centering
%    \includegraphics[width=\linewidth]{asn.png}
%    \caption{Using ASN.1 definitions during modelling.}
%    \label{fig:asn1}
%\end{figure}

However, even if ASN.1 formalises data structures and its encoding/decoding operations, there are no formal specifications of the sanity/security checks to be performed on received data. This makes standards susceptible to ambiguous or misinterpretations that may lead to vulnerabilities in the implementation. Therefore, even though the sample packets provide an overview of the contents of messages, it is still necessary to carefully analyse natural language in standards so that the expected behaviour of the agents towards data contained in messages is captured. Having a well-specified language for expected behaviour and security checks as part of the standardisation process would both mitigate problems related to vulnerabilities in implementations as well as improve the ability to formally verify security properties.
 
% corrresponding language too ASN.1 that specifies checks

\subsection{Model overview}
% 1 page
%DAG of rules (NO)
%Say something about the model size... 974 lines is quite massive

% 22 rules

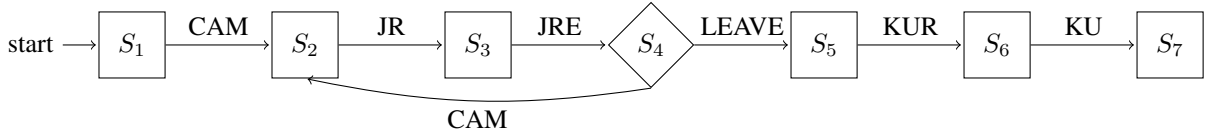
\begin{figure*}[t!]
  \centering
\begin{comment}
\begin{tikzpicture}[shorten >=1pt,node distance=2.3cm,on grid,auto] 
   \node[state,initial] (q_1)  [rectangle,draw] {$S_1$}; 
   \node[state] (q_2) [right=of q_1] [rectangle,draw] {$S_2$};
   \node[state] (q_3) [right=of q_2] [rectangle,draw] {$S_3$}; 
   \node[state] (q_4) [right=of q_3] [rectangle,draw] {$S_4$}; 
   \node[state] (q_5) [right=of q_4] [rectangle,draw] {$S_5$}; 
   \node[state] (q_6) [right=of q_5] [rectangle,draw] {$S_6$}; 
   \node[state] (q_7) [below=of q_6] [rectangle,draw] {$S_7$}; 
   \node[state] (q_8) [left=of q_7] [rectangle,draw] {$S_8$}; 
   \node[state] (q_9) [left=of q_8] [rectangle,draw] {$S_9$}; 
   \node[state] (q_10) [left=of q_9] [rectangle,draw] {$S_{10}$}; 

    \path[->] 
    %(q_0) edge [loop above] node {CAM} ()
    (q_1) edge  node {CAM} (q_2)
    (q_2) edge node {JR} (q_3)
    (q_3) edge node {JRE} (q_4)
    (q_4) edge node {CAM} (q_5)
    (q_5) edge node {JR} (q_6)
    (q_6) edge node {JRE} (q_7)
    (q_6) edge node {JRE} (q_7)
    (q_7) edge node {LEAVE} (q_8)
    (q_8) edge node {KUR} (q_9)
    (q_9) edge node {KU} (q_10)

    ; 
\end{tikzpicture}

\vspace{10px}

\end{comment}

\begin{tikzpicture}[shorten >=1pt,node distance=2.3cm,on grid,auto] 
   \node[state,initial] (q_1) [rectangle,draw]   {$S_1$}; 
   \node[state] (q_2) [right=of q_1] [rectangle,draw] {$S_2$};
   \node[state] (q_3) [right=of q_2] [rectangle,draw] {$S_3$}; 
%   \node[state] (q_4) [right=of q_3] [rectangle,draw] {$S_4$}; 
   \node[state] (q_5) [right=of q_3] [diamond,draw] {$S_4$}; 
   \node[state] (q_6) [right=of q_5] [rectangle,draw] {$S_5$}; 
   \node[state] (q_7) [right=of q_6] [rectangle,draw] {$S_6$};
   \node[state] (q_8) [right=of q_7] [rectangle,draw] {$S_7$};

    \path[->] 
    %(q_0) edge [loop above] node {CAM} ()
    (q_1) edge  node {CAM} (q_2)
    (q_2) edge node {JR} (q_3)
    (q_3) edge node {JRE} (q_5)
    %(q_4) edge node {} (q_5)
    (q_5.south) edge [bend left=10] node {CAM} (q_2.south)
    (q_5) edge node {LEAVE} (q_6)
    (q_6) edge node {KUR} (q_7)
    (q_7) edge node {KU} (q_8)

    ; 
\end{tikzpicture}
    \vspace{-10pt}
  \caption{Simplified diagram of protocol steps (Cooperative Awareness Message (CAM), Join Request (JR), Join Response (JRE), Leave, Key Update Request (KUR), and Key Update (KU)).}
  \label{fig:model-diagram}
\end{figure*} %  placed here so it appears in the protocol model

The security verification of Ensemble was performed through two \tamarin model variants which we define as static and dynamic. Fig.~\ref{fig:model-diagram} depicts the main steps of the protocol operation. Our \textbf{static model} has a fixed sequence of actions from steps $S_1$ to $S_7$ between three agents that includes the announcement of a platoon, a join operation followed by a second join to form a platoon with three members, a leave from the third member and a re-keying process to update the group key. The \textbf{dynamic model} supports the same steps, however with an unlimited number of vehicles that can form an unbounded number of platoons (although with the limitation that each vehicle engages in only one session). This model enables the flexibility of many different scenarios, which include unbounded number of joins, a leave from any of the followers, a key update request to the leader, and propagation of the new group key through a series of key update messages until the interaction ends. We have not analysed the growth of a platoon after a leave has occurred (however, this growth would involve the same interactions from $S_2$ to $S_4$ as shown in Fig.~\ref{fig:model-diagram}).

To support several platoons in the dynamic model, we follow the Ensemble ASN.1 definitions to include a platoon identifier in messages subsequent to the join response, and leverage such identifiers to claim the honesty of vehicles in a certain platoon (i.e., restrict the revealing of keys in the platoon). 

Our models consist of rules that represent the public key infrastructure, initialisation of the vehicles and platoons, sending and receiving messages, and the revealing of keys by the attacker. The rules and properties take approximately 900 lines of text to be defined, and we use \texttt{let} bindings to make the messages more readable and clearer to maintain.

%We model platoons containing up to three vehicles as longer platoons cause state explosion in the verification. While platoons of longer lengths have been deployed on test tracks, there have been no real-world trials on public roads with more than three vehicles. Problems that occur for long platoons include communication issues between head and tail and interaction with other vehicles (e.g., on and off ramps). 

We model certificates as the persistent fact defined below, and it essentially captures the binding between an identity and a public key by a trusted certificate authority (CA with a long-term key $\ltkCA$). In our diagrams, the certificate for a public identity $\idN$ is represented as $\certN$.

\begin{definition}
  A certificate for an identity \idN\ is modelled via the following persistent fact:
  \[
!Cert(\tuple{\idN,\ pk(\ltkN),\ sign(\tuple{\idN,\ pk(\ltkN)},\ \ltkCA)}),
  \]
  where $pk(\ltkN)$ is the public key for $n$ and $sign()$ is the signing operation 
  \label{def:cert}
\end{definition}

\definecolor{gris}{gray}{0.85}
\begin{figure*}[t!]
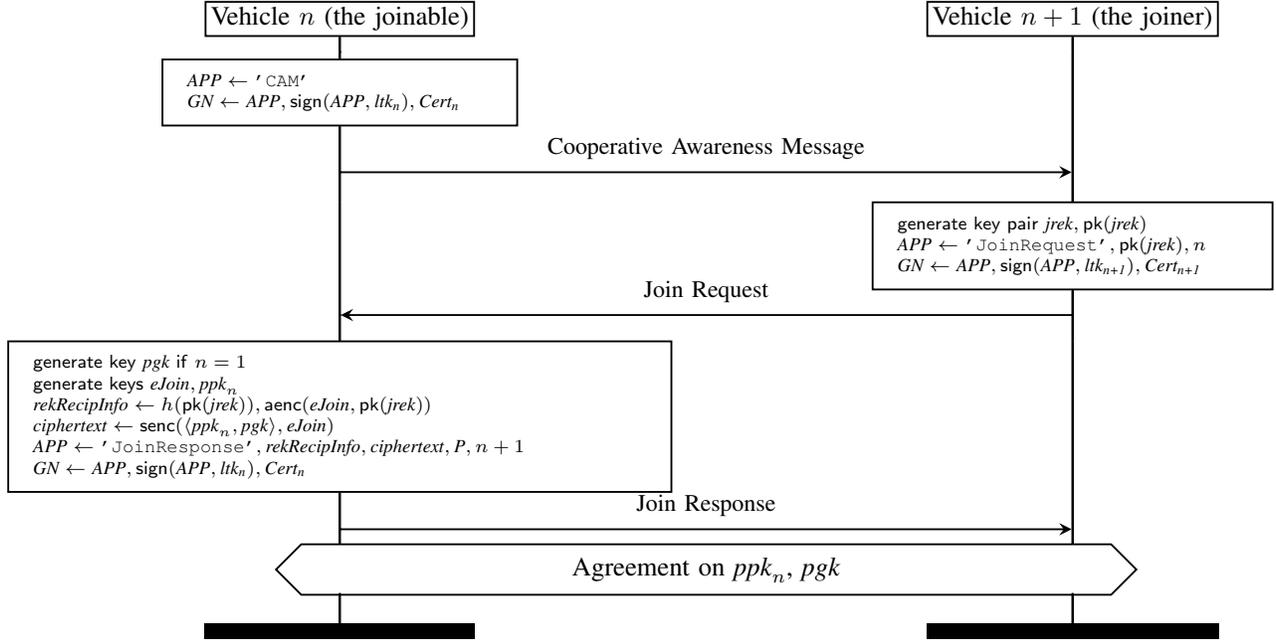

  \centering
  \setmsckeyword{}
  \drawframe{no}    % uncomment to not draw a frame
  \begin{msc}[
    /msc/title top distance=0cm,
    /msc/first level height=.2cm,
    /msc/last level height=0.4cm,
    /msc/head height=0cm,
    /msc/instance width=0cm,
    /msc/head top distance=0.5cm,
    /msc/foot distance=-0.0cm,
    /msc/instance width=0cm,
    /msc/condition height=0.2cm
    ]
    {}
    %%%%%%%%%%%%%%%%%% CONFIG %%%%%%%%%%%%%%%%%%%%%%%%%
    \setlength{\instwidth}{0\mscunit} % to remove default box below agents 
   \setlength{\instdist}{6cm}  % defalut value between agent
    % \setlength{\regionbarwidth}{0\mscunit}
    % \setlength{\instfootheight}{0\mscunit}
    % \setlength{\instheadheight}{.5\mscunit}
    % \setlength{\bottomfootdist}{0\mscunit}

    %%%%%%%%%%%%%%%%%% AGENTS %%%%%%%%%%%%%%%%%%%%%%%%%
   \declinst{A}{}{Vehicle $n$ (the joinable)}
    \declinst{B}{}{Vehicle $n+1$ (the joiner)}
    
    \nextlevel[0.2]
    
        %%%%%%%%%%%%%%%%%% ACTIONS %%%%%%%%%%%%%%%%%%%%%%%%%
    %%% AUTH TOKEN ---> SN ---> UE
    \action*{\parbox{4.4cm}{\scriptsize{$
          \begin{array}[c]{l}
            \APP\leftarrow \CAM  \\
            \GN\leftarrow \APP, \sign(\APP, \ltkN), \certN 
          \end{array}
          $}}}{A}

\nextlevel[3]
    
    \mess{\messLabel{Cooperative Awareness Message}}{A}{B}
    
    \nextlevel[0.8]
    
    %%% RESPONSE UE --> SN
    \action*{\parbox{5cm}{\scriptsize{$
          \begin{array}[c]{l}
           \mathsf{generate~key~pair}\ \jrek, \pkjrek \\
           \APP\leftarrow \JR, \pkjrek, \idN \\
           \GN\leftarrow \APP, \sign(\APP, \ltkNp), \certNp 
          \end{array}
          $}}}{B}
    \nextlevel[3]
    %\nextlevel[2]
    
    \mess{\messLabel{Join Request}}{B}{A}
    
\nextlevel[0.7]

%  rekRecipInfo = <h(pk(~jrekBA)), aenc(~ephemeralAB, pk(~jrekBA))>
%  ciphertext = <senc(<~ppkAB, ~pgk>, ~ephemeralAB)>
%  app = <'JRE', rekRecipInfo, ciphertext, $B>
%  cert = <$A, pk(~ltkA), sigCA>
%  gn = <app, sign(app, ~ltkA), cert>
  
    \action*{\parbox{8.5cm}{\scriptsize{$
          \begin{array}[c]{l}
            \mathsf{generate~key~} \pgk \mathsf{~if}\ n = 1\\
            \mathsf{generate~keys}\ \ejoin, \ppkN \\
           \rekRecipInfo\leftarrow h(\pkjrek), \aenc(\ejoin, \pkjrek) \\
           \ciphertext\leftarrow \senc(\tuple{\ppkN, \pgk}, \ejoin)\\
           \APP\leftarrow \JRE, \rekRecipInfo, \ciphertext, \platoonID, \idN+1 \\
           \GN\leftarrow \APP, \sign(\APP, \ltkN), \certN 
          \end{array}
          $}}}{A}
    \nextlevel[5]
    
    \mess{\messLabel{Join Response}}{A}{B}

    \nextlevel[0.4]
    \condition{Agreement on $\ppkN$, $\pgk$}{A,B}
    \nextlevel[1.3]

  \end{msc}
  \vspace{-10pt}
  \caption{The modelled Ensemble join operation.}
  \label{fig:join-diagram}
\end{figure*}

We now proceed to explain the protocol model in two steps. First, we show the interaction that occurs when a new vehicle joins the platoon (Join procedure). Then we describe a full run of the protocol with two joins and one leave (Full run). The diagrams consider messages of the more expressive dynamic model, which includes platoon identifiers and vehicle position multisets.

\subsubsection{Join procedure}
The sequence diagram in Fig.~\ref{fig:join-diagram} shows a \texttt{join} operation from Vehicle $n+1$ (the joiner) to Vehicle $n$ (the joinable). The \texttt{join} begins with the joinable sending a CAM to neighbours announcing the availability to join the platoon (or to create one). The CAM is unencrypted at the application layer and follows the signing of messages in the GeoNetworking layer (all messages are signed and carry the vehicle's certificate). 

Whenever the joiner receives a CAM advertising a platoon 
%(in our model we do not make distinctions between different CAMs) 
that it wishes to join, a \texttt{join request} message is prepared. The joiner generates a short-term asymmetric key pair $\jrek,\ pk(\jrek)$ and includes the public key in the request sent to the joinable along with the identity of the target joinable vehicle. 

Once the joinable vehicle $n$ receives the \texttt{join request}, it generates three keys: a platoon participant key ($\ppkN$), a platoon group key (\textit{pgk} in case it is the first join, otherwise re-transmit previously generated), and an ephemeral join key (\textit{eJoin}). Both $\ppkN$ and \textit{pgk} are encrypted with \textit{eJoin}, and \textit{eJoin} itself is encrypted with the public key of \textit{jrek} which was sent by the joiner. Recall from Section~\ref{subsec:standards} the use of \texttt{rekRecipInfo} when an ephemeral key is encrypted with a public key that was not obtained from a certificate.

Note that using the \texttt{rekRecipInfo} is discouraged by the security standards as it may introduce \textit{misbinding} attacks. In Ensemble, the inclusion of an intended receiver in the \texttt{join response} mitigates such risk according to our analysis, however, we show in Section~\ref{sec:misbinding} the possible outcome if an intended receiver is omitted or not checked. The following is a quotation from the IEEE 1609.2 standard:

\quoteIEEE{C.7}{It is therefore recommended that secure data exchange entity designers who use public key encryption make use of either public keys in certificates or public keys in signed secured protocol data units (SPDUs), and avoid “raw” public keys because they do not mitigate this misbinding threat.}

The \texttt{join response} transmits the keys along with the current platoon identifier $P$, the intended receiver, and the platoon position of the joiner. Note that we simplify the intended receiver and platoon position in the diagram denoted as $n+1$, whereas the model uses a public identity for the intended receiver and a multiset for the platoon position as previously discussed.

\subsubsection{Full run}
In a real case, there is no upper bound on the number of steps that can be taken in the Ensemble protocol. Even if there is a limit on the number of platoon members, nodes can keep joining and leaving indefinitely. However, we consider it a full run when all message types have been sent. We now describe a scenario where two nodes join a leader node so that the platoon reaches a length of three. Once the third node has joined the platoon it initiates a leave procedure which causes a key update mechanism. 

%We remind the reader that our dynamic model does not restrict the number of joins and key update messages are generated by every pair of remaining platoon members to propagate the updated group key after a leave occurs.

\begin{figure*}[t]
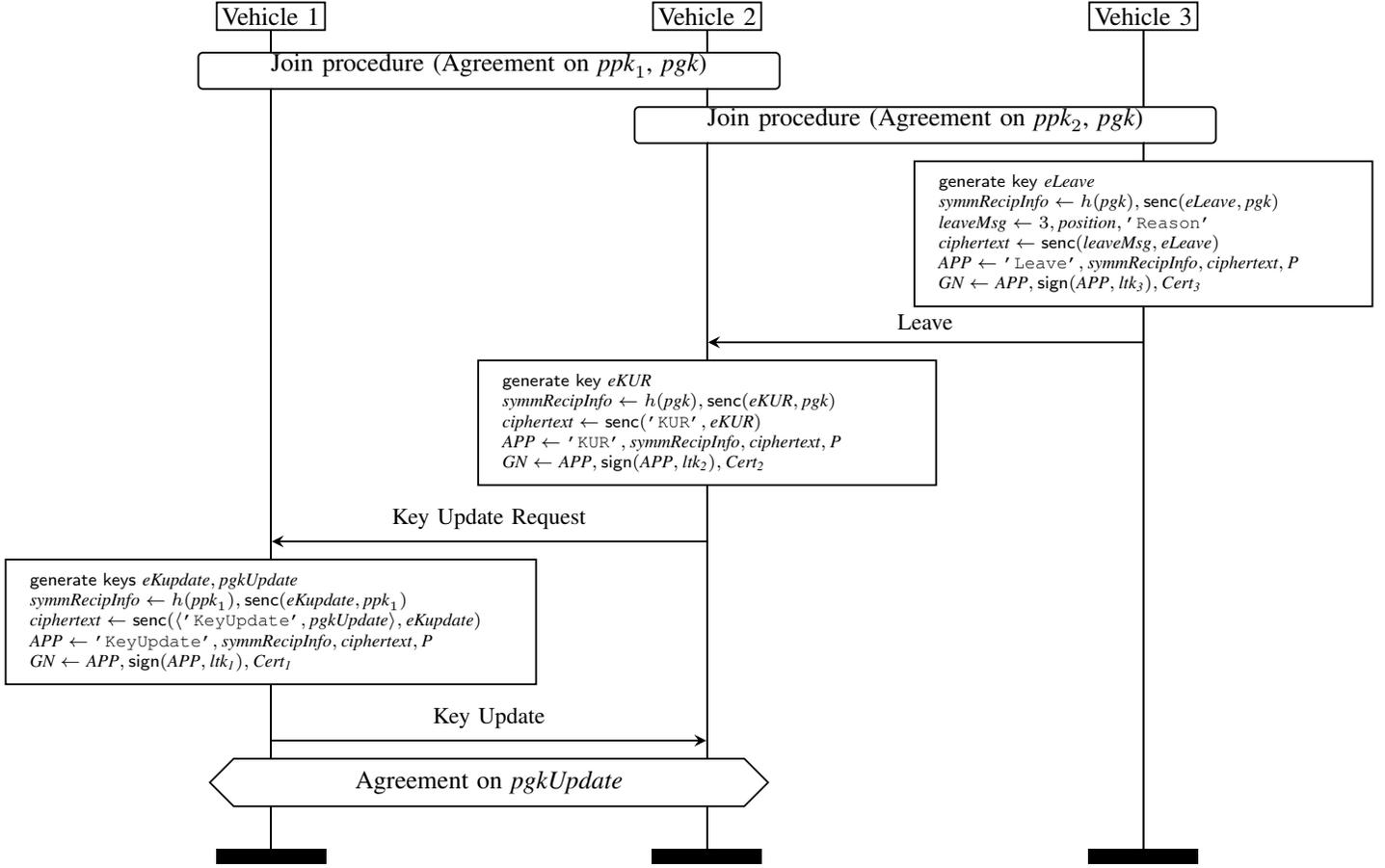

  \centering
  \setmsckeyword{}
  \drawframe{no}    % uncomment to not draw a frame
  \begin{msc}[
    /msc/title top distance=0cm,
    /msc/first level height=.2cm,
    /msc/last level height=0.4cm,
    /msc/head height=0cm,
    /msc/instance width=0cm,
    /msc/head top distance=0.5cm,
    /msc/foot distance=-0.0cm,
    /msc/instance width=0cm,
    /msc/condition height=0.2cm
    ]
    {}
    %%%%%%%%%%%%%%%%%% CONFIG %%%%%%%%%%%%%%%%%%%%%%%%%
    \setlength{\instwidth}{0\mscunit} % to remove default box below agents 
    \setlength{\instdist}{4.5cm}  % defalut value between agent
    % \setlength{\regionbarwidth}{0\mscunit}
    % \setlength{\instfootheight}{0\mscunit}
    % \setlength{\instheadheight}{.5\mscunit}
    % \setlength{\bottomfootdist}{0\mscunit}

    %%%%%%%%%%%%%%%%%% AGENTS %%%%%%%%%%%%%%%%%%%%%%%%%
   \declinst{A}{}{Vehicle 1}
    \declinst{B}{}{Vehicle 2}
    \declinst{C}{}{Vehicle 3}
    
    %\nextlevel[0.5]

%    \mess{\messLabel{Cooperative Awareness Message}}{A}{B}
    
 %   \nextlevel[1]

  %  \mess{\messLabel{Join Request}}{B}{A}
    
   % \nextlevel[1]    
    %\mess{\messLabel{Join Response}}{A}{B}

\nextlevel[0.2]
    
    \referencestart{a}{\vspace{-10mm}Join procedure (Agreement on $\ppkA$, $\pgk$)}{A}{B}
    \nextlevel[1]
    \referenceend{a}
    
    %\condition{Agreement on $\ppkA$, $\pgk$}{A,B}
    %\nextlevel[2]

    %\mess{\messLabel{Cooperative Awareness Message}}{B}{C}
    %\nextlevel[1]

    %\mess{\messLabel{Join Request}}{C}{B}
    %\nextlevel[1]
    
    %\mess{\messLabel{Join Response}}{B}{C}

    \nextlevel[0.5]
   \referencestart{b}{\vspace{-10mm}Join procedure (Agreement on $\ppkB$, $\pgk$)}{B}{C}
    \nextlevel[1]
    \referenceend{b}   
    \nextlevel[0.5]

    %      symmRecipInfo = <h(~pgk), senc(~ephemeralLeave, ~pgk)>
   % leaveMessage = <$C, 'position', 'reason'>
  %  ciphertext = <senc(leaveMessage, ~ephemeralLeave)>
 %   app = <'LEAVE', symmRecipInfo, ciphertext>
%    gn = <app, sign(app, ~ltkC), <$C, pk(~ltkC), sigCA>>

   \action*{\parbox{6cm}{\scriptsize{$
          \begin{array}[c]{l}
            \mathsf{generate~key}\ \eLeave \\
           \symmRecipInfo\leftarrow h(\pgk), \senc(\eLeave, \pgk) \\
           \LEAVEMSG\leftarrow \idC, \pos, \reason \\
           \ciphertext\leftarrow \senc(\LEAVEMSG, \eLeave)\\
           \APP\leftarrow \LEAVE, \symmRecipInfo, \ciphertext, \platoonID\\
           \GN\leftarrow \APP, \sign(\APP, \ltkC), \certC
          \end{array}
          $}}}{C}
          
   \nextlevel[5]
    
    \mess{\messLabel{Leave}}{C}{B}

    \nextlevel[0.5]

    % fresh ephemeralKUR
    %        symmRecipInfo = <h(~pgk), senc(~ephemeralKUR, ~pgk)>
     %   ciphertext = <senc('KUR', ~ephemeralKUR)>
     %   app = <'KUR', symmRecipInfo, ciphertext>
     %   cert = <$B, pk(~ltkB), sigCA>
     %   gn = <app, sign(app, ~ltkB), cert>
    
   \action*{\parbox{6cm}{\scriptsize{$
          \begin{array}[c]{l}
            \mathsf{generate~key}\ \eKUR \\
           \symmRecipInfo\leftarrow h(\pgk), \senc(\eKUR, \pgk) \\
           \ciphertext\leftarrow \senc(\KUR, \eKUR)\\
           \APP\leftarrow \KUR, \symmRecipInfo, \ciphertext, \platoonID\\
           \GN\leftarrow \APP, \sign(\APP, \ltkB), \certB
          \end{array}
          $}}}{B}
   \nextlevel[5]
    
    \mess{\messLabel{Key Update Request}}{B}{A}
    
        \nextlevel[0.5]

        %Fr(~ephemeralKupdate),
        %Fr(~pgkUpdate)
        %    symmRecipInfo = <h(~ppkAB), senc(~ephemeralKupdate, ~ppkAB)>
        %ciphertext = <senc(<'KUpdate', ~pgkUpdate>, ~ephemeralKupdate)>
        %app = <'KUpdate', symmRecipInfo, ciphertext>
        %cert = <$A, pk(~ltkA), sigCA>
        %gn = <app, sign(app, ~ltkA), cert>
    
   \action*{\parbox{7cm}{\scriptsize{$
          \begin{array}[c]{l}
            \mathsf{generate~keys}\ \eKupdate, \pgkUpdate \\
           \symmRecipInfo\leftarrow h(\ppkA), \senc(\eKupdate, \ppkA) \\
           \ciphertext\leftarrow \senc(\tuple{\KUpdate, \pgkUpdate}, \eKupdate)\\
           \APP\leftarrow \KUpdate, \symmRecipInfo, \ciphertext, \platoonID\\
           \GN\leftarrow \APP, \sign(\APP, \ltkA), \certA
          \end{array}
          $}}}{A}
   \nextlevel[5]
    
    \mess{\messLabel{Key Update}}{A}{B}
    
    \nextlevel[0.5]
    \condition{Agreement on $\pgkUpdate$}{A,B}
    \nextlevel[1.7]

  \end{msc}
  \vspace{-10pt}
  \caption{The modelled Ensemble protocol interactions (the join messages from Fig.~\ref{fig:join-diagram} are omitted).}
  \label{fig:protocol-diagram}
\end{figure*}

In Fig.~\ref{fig:protocol-diagram} we simplify the diagram by omitting the \texttt{join} messages. In practice, each join procedure box can be interpreted as an instance of the interactions from Fig.~\ref{fig:join-diagram}. From then on, the sequence diagram represents a \texttt{leave} from Vehicle 3 and a \texttt{key update} procedure so that the remaining members agree on a new platoon group key (which we denote $\pgkUpdate$).

A \texttt{leave} message contains the identity of the leaving vehicle, its position and reason to leave, and is encrypted with an ephemeral leave ($eLeave$) symmetric key. The $eLeave$ key is encrypted with the platoon group key ($pgk$) and is included in a \texttt{symmRecipInfo} recipient data structure. In our model, we represent the position as a multiset and reason as a constant given that no checks are performed on this term.

As soon as Vehicle 2 receives the \texttt{leave} broadcast, it prepares a \texttt{key update request (KUR)} message so that the leader instantiates a new group key. The leader proceeds to generate a $pgkUpdate$ key and includes it in a \texttt{KeyUpdate} message. Both \texttt{KUR} and \texttt{KeyUpdate} messages employ \texttt{symmRecipInfo}, however, while the former uses $pgk$ to encrypt the ephemeral key, the latter uses $ppk_n$ which is private to every pair of adjacent vehicles. Note that in both \texttt{KUR} and \texttt{KeyUpdate} we simply encrypt the message name constants since the contents of these messages are currently under specified in the documentation. The agreement on $pgkUpdate$ is shown at the end of the interaction in the figure.

\definecolor{gris}{gray}{0.85}

%subsection about variants of the model (static and dynamic)
% 2 automata with static and dynamic model representations

\subsection{Verification goals}
\label{sec:verification-goals}
% 0.75 - 1 page
% Honest is a node that we have either sent or received a message

% Honest as soon as hear a message form a node

% Honest @ #l shared pgk ephemeralLeave

The verification goals are divided into \textit{liveness}, \textit{secrecy}, and \textit{authenticity}. Liveness ensures that our model can be executed as expected. See Section~\ref{sec:results} for details of the liveness checks that are performed.

For every key (class of keys in the dynamic model) employed in the life cycle of the protocol, we introduce a secrecy verification lemma based on Definition~\ref{def:prop_secrecy}. There is one important variation in the action facts and lemma construction between the static and dynamic models. Since the static model is composed of one static run, the nodes that participate in the protocol are marked as honest globally. In the dynamic model, since many platoons can be created, honesty claims are done in a per-platoon basis. This allows an attacker to compromise keys of vehicles that are not members of the platoon for which a property is being verified. In the following definitions we consider the use of a platoon identification. Our rules contain \factStyle{Honest(P, n)} action facts that mark an identity $n$ in a platoon $P$ as a benign vehicle (participated in the protocol run and satisfied checks such as signature verification). Therefore, the secrecy formulas define that terms considered secret can not be deduced by an attacker unless it has revealed (through a reveal rule that contains a \factStyle{Rev()} action fact) any subset of keys on which the secret depends. We explain such key dependency relations further in Section~\ref{sec:key_ordering}.

\begin{definition}
Secrecy lemma with platoon identification:
  \begin{equation*}
  \begin{split}
  \forall\ P\ x\ i.\ \factStyle{Secret\_key(P, x)}@i \Rightarrow \\
  (\lnot(\exists\ j.\ \factStyle{KU(x)}@j)\ \\
 |\ (\exists\ n\ k\ r\ h.\ \factStyle{Rev('c', n, k)}@r\ \&\ \factStyle{Honest(P, n)}@h))
 \end{split}
  \end{equation*}
  \label{def:prop_secrecy}
\end{definition}

Intuitively, a secret $x$ instantiated in a platoon with identifier $P$ is either (1) not known by the attacker, or (2) an honest vehicle in the platoon $P$ revealed a secret $k$ of class '\textit{c}' on which $x$ depends. We employ variations of this lemma structure to account for necessary dependencies.

%\begin{definition}
%Secrecy lemma without platoon identification:
%
%  \begin{equation*}
%  \begin{split}
%  \forall\ x\ i.\ \factStyle{Secret\_key(x)}@i \Rightarrow \\
%  (\lnot(\exists\ j.\ \factStyle{KU(x)}@j)\ \\
% |\ (\exists\ n\ r.\ \factStyle{Rev(n)}@r\ \&\ \factStyle{Honest(n)}@i))
% \end{split}
%  \end{equation*}
%  \label{def:prop:secrecy}
%\end{definition}

%Intuitively, this formula states that if $x$ has been marked as a secret key, then either (1) there is no time point $j$ when the attacker knows $x$, or (2) there is some trusted node $n$ which at some time point $r$ revealed its long-term key. Clearly, nodes should not reveal their long-term keys, and the reason for stating the formula this way is to allow for non-trusted nodes to reveal their long-term key and still be able to prove secrecy of certain keys. 

For the authenticity properties we follow Lowe's hierarchy~\cite{596782} and the standard formula definitions according to the \tamarin manual. In the dynamic model, we also consider the platoon identification as part of the agreed data. Namely, we specify \textit{aliveness}, \textit{weak agreement}, and \textit{non-injective agreement} properties. Since our models support a single session per vehicle, we have not analysed injective agreement properties in this work.

To verify these authenticity formulas we annotate the model rules with \factStyle{Running$(m, n, t)$} and \factStyle{Commit$(n, m, t)$} action facts that specify that vehicles $n$ and $m$ agree on their roles and the data represented by $t$. For a full description of the interpretation of these properties we refer to the \tamarin manual. 

%Here, we merely highlight the formula for non-injective agreement for the static case as we describe the use of \factStyle{Commit()} later in our oracle heuristics.

%lemma non_injective_agreement:
%/* when agent 'a' completes the protocol apparently with agent 'b', then 'b' has %previously been running the protocol apparently with 'a'. In addition, they were %both taking their roles in the protocol and agreed on message 't'. */
%  "All a b t #i. Commit(a, b, t) @ #i ==>
%    (Ex #j. Running(b, a, t) @ #j)
%    | (Ex Y #k. Rev(Y) @ #k & Honest(Y) @ #i)"

%\begin{definition}
%  Non-Injective Agreement is modelled via the following formula:
%  \begin{equation*}
%      \begin{split}
%          \forall\ n\ m\ t\ i.\ \factStyle{Commit(n, m, t)}@i \implies
%    \\
%   ((\exists\ j.\ \factStyle{Running(m, n, t)}@j)\\
%   |\ (\exists\ r\ k.\ \factStyle{Rev(r)}@k\ \&\ \factStyle{Honest(r)}@i))
%      \end{split}
%  \end{equation*}
%
%  \label{def:prop:secrecy}
%\end{definition}

%Intuitively this property states that (unless the long-term key of a presumably honest node has been revealed) if a vehicle $n$ commits to having had an interaction with node $m$ in their respective roles and regarding some fact $t$, then $m$ also believes it has interacted with $n$ and they both agree on $t$.

\subsection{Assumptions}
% 0.5 page

% replay
% random number

In this section we describe assumptions related to the replay protection of messages, random numbers, and pattern matching used in the model.

In order to determine whether a vehicle should accept replayed messages, we have identified several places where this concept is mentioned in the relevant documents. The GeoNetworking layer could potentially be used to reject replayed messages. However, all Ensemble messages are single-hop packets which do not carry sequence numbers, so this cannot be used to prevent message replay, and \ETSISEC does not discuss replays. The \IEEESTD standard describes a mechanism that can be used to avoid message replay attacks since it states that identical signed messages are not accepted for a given predetermined time as follows. 

%In GeoNetworking, replay protection is provided when a sequence number is available in the message according to the following quotation.

%\quoteGN{A.1}{The GeoNetworking protocol applies the sequence number-based method for duplicate packet detection (clause A.2) to multi-hop packets (GUC, TSB, GBC, GAC, LS Request, LS Reply). This method is not applied to GeoNetworking single-hop packets (BEACON and SHB) since these packet types do not carry a SN field.}

%Altough GeoNetworking could potentially be used to reject replayed messages, all Ensemble messages are single-hop packets which do not carry sequence numbers. We refer to the following quotation to support this claim.

%\quoteEnComm{3.5.3}{The ENSEMBLE platooning protocol will use the single-hop broadcast GN messages to distribute CAMs, platoon control as well as platoon management messages such as join/leave request/response. Although GN unicast messages could have been selected for management messages, the GN unicast message header imposes unnecessary overhead and complexity.}

%Although \ETSISEC does not clarify aspects regarding replayed  messages, \IEEESTD provides a mechanism to thwart that. In particular, it states that identical signed messages are not accepted for a given predetermined time as follows. 

\quoteIEEE{5.2.4.2.6}{The replay detection service indicates that a signed SPDU is a replay if the entire encoded signed SPDU, including signature and other fields such as generation time inserted by the secure data service, is identical to a recently received SPDU.} 

Ensemble does not explicitly specify that the replay protection from \IEEESTD must be enforced. Having optional security mechanisms is clearly a potential weakness since it delegates important security aspects to choices made in the implementation stage. Still, we believe that the most reasonable assumption according to this statement is that message replay is prevented by the protocol (valid re-transmissions of messages can be done by updating the timestamp, and agents check for uniqueness given a recentness parameter).
%Replay protection is requested as a parameter to a signed data verification function \textit{Sec-SignedDataVerification.request} from \IEEESTD.

% check documents for DATA CHECK (signatures)

We formalise this replay protection as a restriction of traces (only consider those that satisfy the restriction formula) by annotating every message reception with \factStyle{Message}$(x, n)$ where $x$ is a signed message and $n$ is the identity of the receiver.

\begin{definition}
  Replay protection is modelled via the following restriction formula:
  \[
 \forall\ x\ n\ i\ j.\ \factStyle{Message(x, n)}@i\ \&\ \factStyle{Message(x, n)}@j\ \Rightarrow\ i = j
 \]

  \label{def:replay_restriction}
\end{definition}

We include this restriction for completeness only since we have not identified any specific attack that could be launched if the restriction is not present (nor ruled it out). Such an analysis would require a model that allows entities to be present in multiple runs and is therefore out of scope for this paper.

%restriction no_replay_to_same_node:
  %"All x A #i #j. Message(x, A) @ #i & Message(x, A) @ #j ==> #i = #j"

Another assumption in the model is that vehicles verify the signature with the corresponding public key of the sender that was included in the certificate, and match the identity in the certificate with the identity stored in a state fact of that run of the protocol.

Moreover, we assume that the certificate authority is trustworthy, i.e., the attacker is unable to compromise its long-term key $\ltkCA$. In practice, one of the possibilities for the attacker would be to forge certificates with arbitrary identities and public keys in order to conduct identity theft of other vehicles. 

Finally, our model assumes that fresh terms (for instance, ephemeral keys) are unique across all runs of the protocol. In addition, when receiving messages, we employ pattern matching instead of deconstruction. Deconstruction explicitly decomposes the terms by applying equations, selecting specific terms from tuples, and performing sanity checks on decomposed terms. Because of this, pattern matching implicitly checks for message formats and expected data types which must be done explicitly in real software.
% 2-2.5 pages

\section{Proof strategy}
\label{sec:proof_strategy}
In this section we define secrecy and authenticity relations between keys, and explain how a partial order of these relations is employed in our proof strategy. In addition, we provide details of the goal prioritisation of the oracle that guides the constraint solver.

% add key parser description

\subsection{Key ordering}
\label{sec:key_ordering}
% 1 page

The combination of keys defined in Ensemble, the public key infrastructure, and ephemeral keys used in message profiles from the security standards considerably increases the complexity of our analysis. Our strategy towards making the analysis tractable is to define the relations between the keys and break the complexity into smaller parts that can be combined to prove the security properties. We first present how this can be done for a static case where all keys are known at design time, and then discuss the extension to the dynamic case where we know the classes of keys.

Let $\mathcal{K}$ be the set of symmetric and asymmetric keys. We define a secrecy dependency relation $\rightarrow\,\subseteq \mathcal{K} \times \mathcal{K}$ such that for two keys $k_A, k_B \in \mathcal{K}$, $k_A \rightarrow k_B$ holds if revealing the key $k_B$ allows the attacker to learn $k_A$.
%determines that the secrecy of $k_A$ depends on the secrecy of $k_B$, that is, the former can be deduced by the attacker whenever it knows the latter. 
We consider that $k_A \rightarrow k_B$ whenever $\mathsf{senc}(k_A, k_B)$ or $\mathsf{aenc}(k_A, \mathsf{pk}(k_B))$ occurs in a message sent over the network (rule 1). 
%Clearly the secrecy dependency relation is transitive and reflexive (follows from the nature of secrets). 
We note that the secrecy dependency relation is reflexive (i.e., $k \rightarrow k$ for all keys $k$ since revealing a key means that the attacker knows it). Moreover, under the assumption that revealing/compromising a key is a stateless operation (i.e., it does not otherwise change any state in the system), then the secrecy dependency relation is also transitive. This means that if $k_A \rightarrow k_B$ and $k_B \rightarrow k_C$, then $k_A \rightarrow k_C$. 
In most applications this relation is also anti-symmetric, thus giving rise to a partial order of keys. 
To define our authenticity relation, we consider \textit{compromising} a term $x$ as either revealing it or being able to generate a $x'$ that will be accepted by other nodes as $x$. We define an authenticity dependency relation $\dashrightarrow\ \subseteq \mathcal{K} \times \mathcal{K}$ such that for two keys $k_A, k_B \in \mathcal{K}$, $k_A \dashrightarrow k_B$ holds if compromising $k_B$ allows the attacker to create another key $k'_A$ that will be accepted by the other nodes as the legitimate $k_A$, which thereby becomes compromised (rule 2). For instance, if node $n$ generates a fresh term $f$ and signs it with its long-term key $ltk_n$, then $f \dashrightarrow ltk_n$. 
%defines that $key_A$ may be chosen arbitrarily by an attacker whenever it knows $key_B$.
The authenticity dependency relation is irreflexive (knowing a long-term key does not allow creating a new long-term key), transitive (proof in appendix), and should be anti-symmetric since otherwise the protocol has a cyclic authenticity dependency. 

In Ensemble, the two key relations $\rightarrow$ and $\dashrightarrow$ are both anti-symmetric (there are no cases where two different keys depend on each other). By taking the union of the two relations (a relation is a set of pairs, so the union of two relations is the aggregation of all pairs from both relations) we arrive at a third relation  whose transitive closure forms a partial order $\rightsquigarrow \subseteq \mathcal{K} \times\mathcal{K}$. Intuitively, whenever $k_A \rightsquigarrow k_B$, compromising $k_B$ will allow the attacker to compromise $k_A$, either directly, or through a chain of learned/replaced keys in which the attacker appears as the legitimate entity that controlled $k_B$. Note that the joint dependency relation can in some cases be automatically deduced from a formal description of a protocol through rules 1 and 2 as we describe in the next subsection.

The dependency relation we have described here assumes a static set of keys, and also that the relation itself is time-invariant. In reality, there are several situations where these assumptions do not hold. In the case of Ensemble, the static model can be immediately analysed as all keys are known a-priori. For the dynamic model on the other hand, we have to consider \emph{classes} of keys by, for example, considering all \textit{ppk} keys as if they were a single key. This means that if \emph{some} \textit{ppk} key depends on another key $k$, then the class of \textit{ppk} will depend on the class of $k$.
Another situation where the assumptions do not hold is if the reveal/compromise can be limited to happen only a finite number of times. In this case, transitivity is not guaranteed to hold so compromising a top-level key does not necessarily mean that all keys "under" also become compromised. Finally, the dependency relation does not specify when a key can be compromised, so it does not account for perfect forward secrecy formulations. Taken together, the dependency graph that we consider should be seen as an abstraction in which a dependency actually means a \emph{possible} dependency. Since the purpose of the ordering relation is to guide the prover on which lemmas/goals to prioritise, having spurious dependencies does not cause erroneous results, but can potentially reduce its usefulness. An extension of our approach would be to let the prover maintain a dynamic key hierarchy at runtime which would at least account for a changing set of keys. We discuss this further in the future work section.

%This results in an abstraction of the real dependency graph that is pessimistic, but since the purpose of the graph is to guide the prover, it does not in itself 

%the reveal/compromise can be limited to happen only a finite number of times. In this situation, transitivity is not guaranteed to hold so compromising a top-level key does not necessarily mean that all keys "under" also become compromised. This would be a desirable security property, but in many cases difficult to implement.

\subsection{Key dependency extractor}
\label{sec:dep_extractor}
%(Start new section)
%Explain parser
% difficult with cycles, parser currently does not work with models that use multisets or XOR

The dependencies formalised through rules 1 and 2 can be automatically extracted from a \tamarin model to support oracle and reusable lemmas construction. To implement this extractor we have extended the \textit{Tamarin to alice\&bob translator}~\cite{kozmaiconverting} in order to parse the model, extract key dependencies, identify term equivalences (for instance, keys with different names across distinct rules) through unification, and then grouping equivalent keys to output a graph of the hierarchy. The process of extracting the key dependency is summarised as follows:

\begin{enumerate}
    \item The model is parsed to instantiate an internal representation~\cite{kozmaiconverting} and an empty directed acyclic graph (DAG) of keys is instantiated.
    \item Terms are identified in each multiset rule of the \tamarin model and subsequently added as nodes to the DAG of keys. Relations are added as edges according to rules 1 and 2 defined in the previous subsection (duplicates may be merged through unification in later steps).
    \item Instantiate empty premise and conclusion lists. Premise and conclusion facts are added to the premise and conclusion lists, respectively.
    \item Facts from the premise and conclusion lists with same name and arity (as well as inputs and outputs from the network) are unified if possible (we use maude for this~\cite{CLAVEL2002187}). This results in a set of term substitutions (unifiers) which we use to find equivalent keys.
    \item A new DAG of equivalent keys is instantiated given the prior DAG of keys and the resulting sets of equivalent keys. Its topological sort represents the key ordering.
\end{enumerate}

%the key dependency extractor works on existing protocols such as Privacy-Preserving OpenID Connect~\cite{10.1145/3320269.3384724}. 

\begin{figure}[h]
    \centering
    \includegraphics[width=\linewidth]{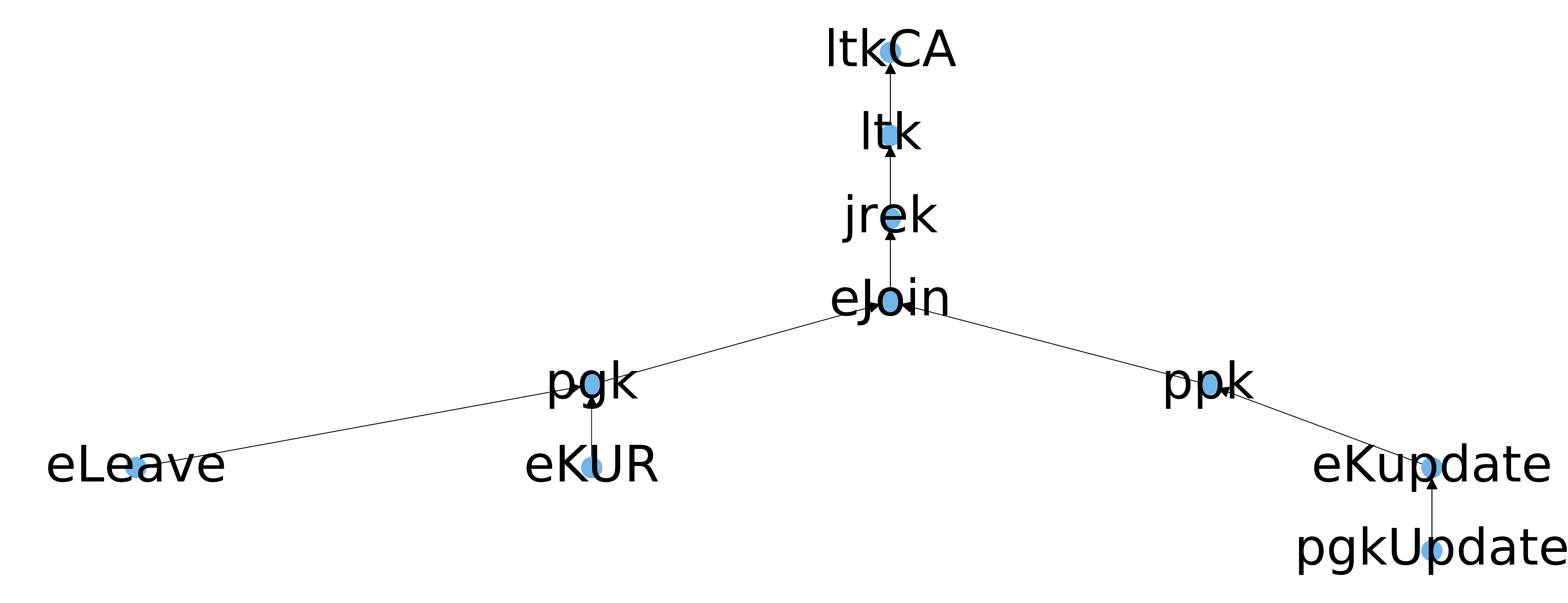}
    \caption{Key dependency graph automatically extracted from our dynamic model.}
    \label{fig:hierarchy}
\end{figure}

We illustrate the partial order of keys in the Ensemble protocol extracted from the dynamic model using our tool in Fig.~\ref{fig:hierarchy}. We see that at the top of this partial order is the long-term key of the certificate authority, and at the bottom are the ephemeral keys as well as the $\pgkUpdate$ key. Edges that are covered by the transitivity of the relations have been omitted in the graph.

Proving security properties involving keys at the bottom of this order requires that one is first able to prove the secrecy of the keys in the upper layers. More concretely, if $k_A \rightsquigarrow k_B$, then any security property (secrecy or authenticity) that depends on $k_A$ also depends on $k_B$. Thus, one should first show the secrecy of $k_B$. In a simple protocol with few keys, this order matters little, but the more complex the key relationships become, the more important it is that the verifier is aware of the key ordering. Often this is implicit in the helper lemmas or oracle instrumentation made for each security property. 

Due to limitations in the model parser, we remove union operations (enabled by the multiset built-in theory) from \tamarin models and other theories are not currently supported (such as XOR and Diffie-Hellman). The relations for such theories could be derived from the message deconstruction rules used in \tamarin, but are out of the scope of our current analysis. In addition, we note that some models might generate cyclic graphs which are currently not supported. %In this case, a maximum branching algorithm could be employed but we leave such investigation to future work.

Despite these limitations, we have employed our proof-of-concept tool in the analysis of a recent work in 5G handover protocols~\cite{10.1145/3448300.3467823}. These models also present a complex relation on keys, and required the modellers to consider the dependencies during the specification of lemmas and oracles. For instance, given that the model allows the revealing of some keys, the secrecy lemmas must restrict the revealing of dependencies. Furthermore, the oracles also seem to consider some dependencies in the prioritisation, similar to what we present in this work. While we have not enhanced the proving efficiency of these models that had already been carefully optimised by experts, we argue that applying the strategies described in this work could be useful during the analysis and modelling of such protocols. Examples of the extracted graphs and further discussions are included in Appendix~\ref{app:handover}.

%\input{key_hierarchy}

%Our hypothesis in this work is that by making the proof strategy order-aware, we can construct a generic oracle for all lemmas, and then just tell it what the order relation looks like so that it prioritises the most relevant branches based on that information.

%Proving secrecy of a key in this partial order can only be done if secrecy of ... {\bf argument for why this order should guide the proving process, and why an order-agnostic prover will not succeed}

\subsection{Inductive helper lemmas}
The possibility to instantiate infinitely many vehicles and platoons associated to the fact that each platoon can grow indefinitely aggregates further complexity to our dynamic model. The model therefore enables loops that result in non-termination when using standard backwards search in \tamarin. To handle such behaviour, \tamarin allows the specification of inductive lemmas which we employ as intermediate helpers.

Recently, Cremers et al. have analysed IEEE 802.11’s WPA2 protocol~\cite{cremers2020formal}, which also contains complex state machines with loops and evolving states. In their work, they specify Wellfoundedness, Uniqueness, and Ordering lemmas. We follow their approach in the creation of such inductive intermediate helper lemmas, and in conjunction employ our key hierarchy prioritisation.

\subsection{Oracle strategy}
\label{sec:oracle}
%oracle 
%helper lemma ordering

We now describe how the use of the linear extension of the partial order of keys is used to guide the solver. Recall from Section~\ref{sec:protocol-sec-verif} that \tamarin checks possible sources for constraints to generate traces that will be used to prove or find a counterexample for a given property. The choice of which constraint to solve (goal) at a given step can be tailored by using \textit{oracles}. To perform the Ensemble verification, we structured the lemmas and developed an oracle so that they leverage the key ordering from our key dependency extractor.

%Proof goals $\mathcal{G}$; Ordered list of message types $\mathcal{M}$; Linearized key ordering $\mathcal{K^\rightsquigarrow}$\\
% sentence that summarizes the checks (which ones are generic, specific to Ensemble, specific to authentication/secrecy)

Due to space constraints, we only present the main prioritisation activities performed by the oracle, which leverage the selection of reusable lemmas to avoid unnecessary case distinctions during proving.

%While our static and dynamic variants are used to model the same Ensemble protocol, they contain several peculiarities that required distinct custom oracles. We present the main approaches that are common to both variants, and the reasoning about the prioritisation of goals and use of key ordering is detailed below. Due to space limitation, we briefly describe the most important points for our strategy.
%Most of the priorities described are of a generic nature that could be applied to any protocol that uses symmetric and asymmetric encryption. Priority number 4 is protocol specific, due to a particularity on how keys are sent in Ensemble. Priorities 7 and 10 are the ones that make this oracle order-aware, thus integrating the information contained in the partial order of keys. Priority number 6 has to do with message order, and thus also provides a way in which the model structure guides the proof strategy.

%\begin{enumerate}
    %\item 
    \textbf{Ordered helper lemmas iff a knowledge goal for the corresponding key exists in the constraint system:} 
    In order to create contradictions earlier, the helpers are prioritised if there is currently a goal for an attacker knowledge of the corresponding key $k$ as \factStyle{KU}$(k)$.  Algorithm~\ref{alg:helper} performs this prioritisation. The algorithm runs for each $k$ according to the ordering of keys.
    %pseudocode
    \begin{algorithm}[h]
 \KwData{Proof goals $\mathcal{G}$; Linearised key ordering $\mathcal{K^\rightsquigarrow}$}
 \KwResult{Ordered list of goals $\mathcal{G^\rightsquigarrow}$}
    \ForEach{$k \in \mathcal{K^\rightsquigarrow}$}{
    \ForEach{$g$ in $\mathcal{G}$}{
    \If(){$g$ is a helper lemma for $k$ and $KU(k) \in \mathcal{G}$}{
      add $g$ to $\mathcal{G^\rightsquigarrow}$
        
    }
    }
    }

 \caption{Oracle priority pseudocode for helper lemmas}
 \label{alg:helper}
 
\end{algorithm}
    %\item \textbf{State/PKI/initialisation premise facts:} State facts model the memory state of a vehicle. By prioritising state fact goals, infeasible traces can be ruled out. Initialisation and PKI facts are also prioritised given that they are required for the execution of the protocol.
    %\item \textbf{Honesty action facts:} Recall from Section~\ref{sec:verification-goals} that lemmas are constructed to prove that our properties hold \textit{or} there was a reveal of the long-term key of a vehicle that was considered honest at a certain time point. Therefore, invalid traces are excluded where an honest node has revealed its long-term key.
    %\item \textbf{Long-term key persistent facts:} These persistent facts hold the map between identities and long-term keys, and are used for generating signatures.
    %\  item \textbf{Symmetric encryption of incorrect pairs:} We have observed that \tamarin attempts to deduce any term from a message that employs the symmetric encryption of two fresh terms with another fresh term, which occurs at the join response in Ensemble according to Fig.~\ref{fig:join-diagram}. Therefore, prioritising this goal usually results in contradictions faster.
    %\item \textbf{Commit action facts:} Commit action facts stem from authenticity properties, cf. Section~\ref{sec:verification-goals}.
    %\item 

    \textbf{Signature of protocol messages:} Following our hierarchical key approach, we introduce helper lemmas that prove that the attacker cannot obtain any long-term key unless it performs a reveal of those keys. Since the authenticity of messages depend on the secrecy of the respective long-term keys (because of the signatures), we prioritise these goals to determine that the attacker is not able to forge signatures or act on behalf of an honest node.
   % \item \textbf{Knowledge of long-term keys:} Long-term keys are only used to generate signatures, hence we prioritise over symmetric and asymmetric encryption.
    %\item \textbf{Symmetric and asymmetric encryption:} Instances of $senc(\_,\_)$ or $aenc(\_,\_)$ will possibly generate \factStyle{KU()} goals that can be then prioritised.
    %\item \textbf{Ordered key knowledge:} Finally, the \factStyle{KU()} goals are prioritised according to the linear extension of our partial order.
%\end{enumerate}

%We have developed the oracle generator with the intention of it being as generic as possible. Still, the internal prioritisation might not be optimal in all cases. In addition to applying it on the Ensemble protocol, we have tested it with the Privacy-Preserving OpenID Connect~\cite{10.1145/3320269.3384724} model, however, it did not show an efficiency gain over the state-of-the-art strategies. We believe that further research on how the model structure can help guide the proof strategy is warranted. 

%\subsection{Other proof heuristics}
% 1 page

%other helper lemmas
%restrictions
%pattern matching
%restricting agents 

%lemma Helper_ltkCA_ind[use_induction, reuse]:
%"All x #i. Helper_ltkCA(x) @#i ==> Ex #j. Secret_ltkCA(x) @#j  &  #j < #i"

%lemma Helper_ltkCA[use_induction, reuse]:
%"All x #i #J. Secret_ltkCA(x) @#i & KU(x) @#J ==> F"

%lemma Helper_ltkA[reuse]:
%"All x #i #J. Helper_ltkA(x) @#i & KU(x) @#J ==> Ex #k. Rev_Ltk(x) @ #k & #k < #J"

%lemma Secret_jrekBA[reuse]:
%  "All x #i. Secret_jrekBA(x) @i ==> not (Ex #j. KU(x) @ #j)  | (Ex Y #k. Rev(Y) @ #k & Honest(Y) @ #i)"

% 1.5 - 2 pages

\section{Results and discussion}
\label{sec:results}
%Introtext (what does this section contain)

% CPU core info
% Intel(R) Xeon(R) Gold 6130 CPU @ 2.10GHz

In this section we present the Ensemble verification results using our \tamarin model variants and their respective proven properties. In addition, an evaluation of our proof strategy is conducted in order to show that leveraging an order-aware oracle is effective.

\subsection{Security verification results}

Considering the security verification goals described in Section~\ref{sec:verification-goals} we prove three kinds of properties, model liveness, secrecy and authenticity. For each of these we describe the resulting security lemmas in the context of our model.

\paragraph{Liveness} To ensure protocol executability in our static variant, we prove that a full run with three vehicles exists. In the dynamic variant we verify that the following is possible: two platoons can be formed with four members each, there exists a leave from a member in a platoon, as well as a key update request and key updates are performed for remaining members.
\paragraph{Secrecy} We prove secrecy of all long-term and short-term keys. We use one lemma per key in the static model as they are instantiated in distinct rules, so there is a total of 15 secrecy lemmas. The dynamic model contains one secrecy lemma for every class of keys (e.g., one lemma proves the secrecy of all platoon participant keys).
\paragraph{Authenticity} We have one lemma each for the authenticity properties aliveness, weak agreement and non-injective agreement (cf. Section~\ref{sec:verification-goals}).

%Altogether there are 20 lemmas that capture the security properties of the platooning protocol. 

%In addition, there are 2 pure helper lemmas, which have been added simply to make the model tractable. 

%By comparing how many of the lemmas that can be proven using the different strategies, we can get a sense of how effective they are.

The properties have been proven for both static and dynamic models. This required making use of all the verification strategies described in Section~\ref{sec:proof_strategy}. 

\subsection{Verification strategy evaluation}
To assess the effectiveness and impact of the proof strategies, we use two experiments: a synthetic protocol generator and a variation of configurations for proving the static model. The experiments are run on a cluster of the Swedish National Supercomputer Centre, where each compute node is equipped with Intel(R) Xeon(R) Gold 6130 CPU @ 2.10GHz with 32 cores and 96~GiB of main memory. 

The synthetic protocol consists of a simple 'ping-pong' protocol in which two nodes communicate and in every interaction instantiate a new symmetric key which is encrypted with the previously received key (the first instance is derived from a pre-shared key). We use the standard \tamarin heuristics and only provide annotations in the model to prioritise certain facts (state, pre-shared key, symmetric encryption, and attacker knowledge of secret). The lemmas are created according to the linear dependency of keys, and the proving is evaluated with and without the reuse of lemmas. In addition, we perform an experiment with the reuse of lemmas that are randomly ordered (for this, the prover is executed ten times with random orders for each key depth).

Each run was granted 8 cores of CPU for 30 minutes and 20~Gib of RAM. Table~\ref{tab:synthetic_results} presents the results. \tamarin is able to automatically prove the secrecy of keys in a depth of 2 in all cases, and up to a depth of 8 when reusing ordered lemmas according to the key hierarchy, which shows how our strategy makes such proofs tractable. In some cases, the random ordering resulted in a possible automatic proof, but took significantly more time to terminate as it was not the optimal order. The key depth of 10 could not be proven and would require further manual intervention (e.g., through an oracle).

\newcommand{\lincheckmark}{\ding{52}}
\newcommand{\lincross}{\ding{54}}

\begin{table}[htb]
    \centering
        \caption{Results of proving the synthetic models with distinct key dependency depth}
    \label{tab:synthetic_results}    
    \begin{tabular}{P{1cm}|P{2cm}|P{2cm}|P{2cm}}
         Key depth & Without reuse (ordered by dependency) & With reuse (random order) &  With reuse (ordered by dependency)\\
         \hline
         \hline
         2 & \lincheckmark & \lincheckmark & \lincheckmark \\
         %\hline
         %\cline{2-3} & Yes & \lincheckmark\\
         \hline
         
         4 & \lincross & 3/10  & \lincheckmark \\
         %\hline
         %\cline{2-3} & Yes & \lincheckmark\\
         \hline
         
         6 & \lincross & 2/10  & \lincheckmark \\
         %\hline
         %\cline{2-3} & Yes & \lincheckmark\\
         \hline
         
          8 & \lincross & \lincross  & \lincheckmark \\
         %\hline
        %\cline{2-3} & Yes & \lincheckmark\\
         \hline
         
          10 & \lincross &  \lincross  & \lincross  \\
         %\hline
         %\cline{2-3} & Yes & \lincross\\
         \hline

         \hline
    \end{tabular}
\end{table}

In our second set of experiments, we run \tamarin on the the static model of Ensemble with several different parameter configurations and measure how many of the security lemmas can be proven with these settings and the computational resources that are used.  %We conduct the evaluation using the static variant of our model given that it currently provides a complete set of messages from Ensemble and our parser is able to extract the key hierarchy and oracle (the dynamic model employs multisets to control the size of the platoon and the parser is currently not able to handle multisets).

We run the prover with four different configurations, outlined as follows. 
\begin{itemize}
    \item {\bf Bare \tamarin} - In this configuration we try to prove the security properties of the protocol without any added proof strategies or helper lemmas.
    \item {\bf Lemma reuse} - Lemmas that for example assert secrecy of keys are set as reusable so that the verifier can assume these lemmas to be true when searching for the proof. 
    \item {\bf Oracle only} - Here we make use of the order-aware oracle but do not reuse lemmas, and must therefore reprove all relevant subresults for every property.
    \item {\bf Order-aware} - In this configuration we use both the order-aware oracle and reuse helper lemmas.
\end{itemize}

The first two configurations should be considered as baselines. The reason for including both the "Oracle only" and "Order-aware" configurations rather than just a single good strategy is to investigate the relative impact of the different aspects of the generated oracle as the ordering of reusable lemmas according to the key hierarchy is an important aspect of its design (see Section~\ref{sec:oracle}).

In addition to the amount of successfully proven lemmas, we measure the resource consumption in terms of computation time and memory usage. Each lemma was run as a separate job in the cluster, and given an allocation of 2 hours on 8 cores and 22~GiB of memory. Jobs that exceeded either the time or memory limit were aborted.

\subsection{Effectiveness of proof strategies}

We now proceed to present the outcome of the second experiments. Table~\ref{tab:provability} shows an overview of how the four different prover configurations performed in terms of proving the 20 different security properties. The Oracle and Reuse columns summarise the key differences between the different configurations (with or without the key-aware oracle, and with or without reusable lemmas). The final three columns show how many of the lemmas that could be proven in the three different categories.

\begin{table}[ht]
    \centering
        \caption{Overview of provability for different proof strategies }
    \label{tab:provability}    
    \begin{tabular}{l|P{0.7cm}|P{0.7cm}|P{0.9cm}|P{0.8cm}|P{1cm}}
         Proof method & Oracle & Reuse & Liveness & Secrecy & Authenticity\\
         \hline
         \hline
         Bare \tamarin & N & N & 1/1 & 0/15 & 0/3\\
         \hline
         Lemma reuse & N & Y & 1/1 & 4/15 & 1/3\\
         \hline
         Oracle only & Y & N & 1/1 & 15/15 & 3/3\\
         \hline
         Order-aware & Y & Y & 1/1 & 15/15 & 3/3\\
         \hline
    \end{tabular}
\end{table}

The results clearly demonstrate the effectiveness of the key-aware oracle, which seems to be the deciding factor to making the model tractable for the verifier. 

%Allowing reuse of lemmas (and thereby ordering of them according to the key hierarchy) does help with the final authenticity lemma which is to prove injective agreement on the pair-wise keys $ppk_1$ and $ppk_2$ as used in a three-vehicle setting. In all the cases where the verification did not finish the reason was that the process ran out of available memory (22~GiB). For the lemmas that were proven the memory consumption was below 600~MiB except in one case (the same lemma as discussed above) which required 2~GiB.

\begin{figure}
    \centering
    \includegraphics[width=\columnwidth]{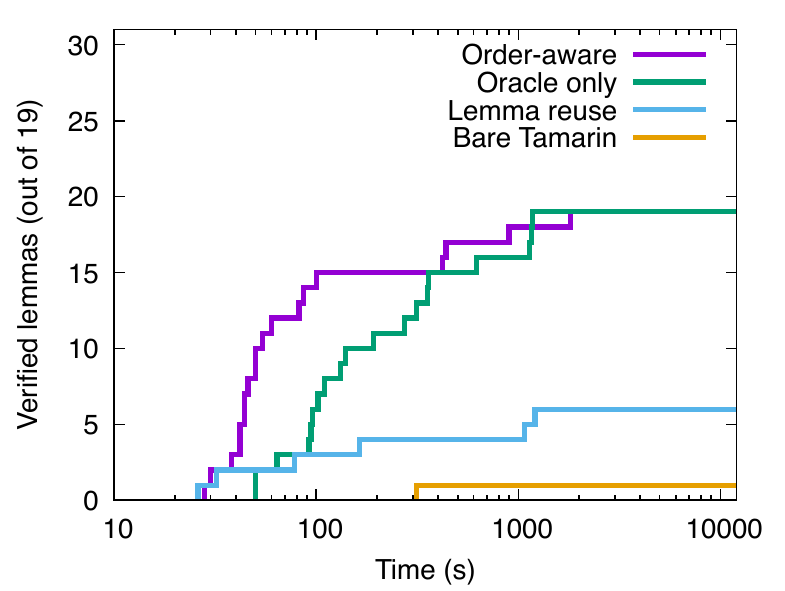}
    \caption{The number of lemmas proven within a given time for the different proof strategies (logarithmic x axis).} %  Note that each lemma is verified in a separate process with its own computational resources, so they do not interfere with each other.
    \label{fig:time}
\end{figure}

Another perspective on the performance of the different strategies is shown in Fig.~\ref{fig:time}. The graph shows time on the X axis (logarithmic scale) and the number of lemmas proven within this time on the Y axis. There is a significant (and expected) performance difference observed when making use of, and ordering, previously proven lemmas. In particular, the fastest 15 lemmas were verified in 86 seconds by the order-aware strategy whereas it took over 10 minutes when lemmas were not reused (Oracle only). 

%We see that the last lemma that was proven by the order-aware oracle with the help of reusable lemmas took almost 2 hours, but that the other 18 lemmas could be verified in around 30 minutes for the two best strategies. Finally, there is a significant (and expected) performance difference observed when making use of, and ordering, previously proven lemmas. In particular, the fastest 15 lemmas were verified in 86 seconds by the order-aware strategy whereas it took over 10 minutes when lemmas were not reused (Oracle only). Interestingly, there were also a couple of cases where the reverse is true which is likely because some lemmas cause the prover engine to explore branches that do not lead anywhere. 

%Time-graph

%memory-graph

%Say something about time-outs vs out-of-memory

% table with properties, good/bad model (intended receiver), result
%weakness in the specification (intended receiver, and join request)

% 1.5 pages

%\begin{itemize}
%%    \item G1. Model executability
%    \item G2. Secrecy of long-
%\end{itemize}

\subsection{Identity misbinding attack}
\label{sec:misbinding}
% 0.5 - 1 page
%Consistency. If two honest parties establish a common session key then both
%need to have a consistent view of who the peers to the session are. Namely,
%if a party A establishes a key K and believes the peer to the exchange to be
%B, then if B establishes the session key K then it needs to believe that the
%peer to the exchange is A; and vice-versa.

An \textit{identity misbinding attack}~\cite{10.1007/978-3-540-45146-4_24, diffie1992authentication}, also referred to as an \textit{unknown key-share attack}~\cite{kaliski2001unknown, 10.1007/3-540-49162-7_12} occurs when two honest parties establish a common session key without a consistent view of each other's identities. In IEEE 1609.2, the use of \texttt{rekRecipInfo} can possibly create a vulnerability to misbinding since the public key is not bound to an identity. In Ensemble, even though this data structure is used in a \textit{Join Response}, misbinding can be mitigated because the identity of the intended receiver is included in the application payload. However, it is not stated explicitly that this information should be validated. The receiver must check that the intended receiver included in the message matches its own identity. In an implementation of the protocol where this is not done, some security properties will be violated. We ran the verification with a modified variant of the static model that captures this aspect. The results can be seen in Table~\ref{tab:intended_receiver} where we see that the weak and non-injective agreement authenticity properties are not satisfied. 

\begin{table}[]
    \centering
        \caption{Authentication failure in the absence of intended receiver}
    \label{tab:intended_receiver}
    \begin{tabular}{P{2.5cm} | P{1.5cm} | P{2cm}}
         Security property & Intended receiver & No intended receiver \\
         \hline
                  \hline
         Aliveness & \lincheckmark & \lincheckmark \\
                  \hline
         Weak agreement & \lincheckmark & \lincross \\
                  \hline
         Non-injective agreement & \lincheckmark & \lincross   \\
                  \hline
    \end{tabular}
\vspace{-4mm}
\end{table}

The possible attacker behaviour is described as follows (see also  Fig.~\ref{fig:join-diagram}). Provided that the check of intended receiver is absent, an inside attacker (who has a valid long-term key and certificate) could replay the CAM (from Vehicle 1) that advertises a joinable platoon to another Vehicle 2, which will send a \texttt{join request}. The attacker extracts the public key of \textit{jrek} from that message and uses it in a join request signed with his own long-term key. Vehicle 1 will send a \texttt{join response} to the attacker, which will transmit it back to Vehicle 2. At the end of the procedure, Vehicle 1 believes that the attacker has joined, whereas Vehicle 2 believes it has joined Vehicle 1, and both share the participant and group keys (note that the attacker can not compromise the secrecy).

% 2-2.5 pages

\section{Related work}
\label{sec:related_work}

Vehicular network security standardisation and its formal analysis is rather recent. Whitefield et al.~\cite{10.1007/978-3-319-68063-7_10} analyse V2X certificate revocation of malicious or misbehaving vehicles with the \rewire scheme using \tamarin. In their analysis, they are able to identify an authentication weakness and propose an extension to mitigate it. Li et al. propose a lightweight privacy-preserving authentication
protocol that is verified with BAN logic and \proverif~\cite{9099433}.

In mobile networks, Basin et al.~\cite{10.1145/3243734.3243846} formalise the 5G authentication and key agreement protocol, and verify security properties using \tamarin. The authors found in their analysis that security goals and assumptions were under-specified or missing. We show a similar situation in our analysis of standards in the vehicular domain. While data structures are often well defined, under-specification of data checks and behaviour can lead to misinterpretation and potential security vulnerabilities.

%Focardi and Luccio~\cite{9155194} propose a symbolic mechanized model for verifying secrecy and authentication in the context of physical unclonable functions. 

With respect to verification and solving theory, Cremers and Mauw~\cite{10.1007/978-3-540-31810-1_12} employ partial order reduction to lower the number of traversed states in checking secrecy of terms in a cryptographic protocol in their tool \scyther. They build on the fact that exchanging two events in a trace might result in equivalent traces with respect to the verified property. In our work, we explore the fact that solving for the knowledge of some terms might not be relevant, and that solving for the knowledge of some terms before others is more efficient.

Schmidt et al.~\cite{6956564} develop an algorithm to verify protocol group key agreement protocols that can handle Diffie-Hellman exponentiation, bilinear pairing, and AC-operators. In their work they extend the operators set and provide constraint reduction rules in \tamarin to support them. They argue for the analysis of dynamic join and leave operations in group protocols, which is also present in our model.

%Dreier et al.~\cite{8429317} advance the XOR model in \tamarin to handle its inherent associativity and commutativity properties and mitigating non-termination of verification of protocols that employ XOR operations.

% 1 - 1.5 pages

\section{Conclusion and Future Work}
\label{sec:conclusion}

% ASN.1 helps modeling of messages

% translation from ASN.1 to MSR

% 

%notation to represent rules for checking data transmitted in messages

We have formally analysed the security of Ensemble, a protocol for vehicular group formation with key establishment and distribution which is currently in pre-standardisation. To conduct the verification, we define secrecy and authentication relations that are applied in a proof strategy based on their partial order. We automate the key hierarchy extraction from our \tamarin models and create oracles to guide the prover based on the ordering of keys. To refine the model of the protocol messages we use ASN.1 definitions from standards and a compiler to generate sample packets, which was useful to avoid misinterpretations or ambiguities from multiple documents. Through our assessment of vehicular network security standards by IEEE and ETSI, we show that although they provide solid security message formats, the implementations may still be susceptible to weaknesses if the expected agent behaviour is not enforced. We show an example of such a weakness in the form of a misbinding attack when appropriate checks are not performed by the vehicles. An interesting point for discussion in the context of standardisation work lies in formally describing agent behaviour towards received data and appropriate security checks.

\tamarin enables the formal analysis of several complex protocols, and may require manual tuning in some cases.  We believe that, ideally, an automated security analysis should be able to derive, without manual intervention, the set of conditions for each cryptographic term in a protocol to remain secret and provide the corresponding proof (currently, modellers must identify such conditions and specify them in the lemmas). Our work to automatically extract key dependencies is a step towards this long-term goal, and many interesting challenges remain. An integration of the dependency analysis in \tamarin at runtime (during proving) would allow much richer reasoning. For instance, this could allow the possibility to consider time and properties that involve forward secrecy. In addition, the extension of dependency relations to account for XOR, Multiset, Diffie-Hellman, and other equational theories are required to support a large class of models. We consider these challenges to be important contributions in future work.

%Other interesting direction for future work include 
%- Integration of key extraction in Tamarin
%- Reasoning with regards to time (PFS)
%- Extending the dependency relations to support Diffie Hellman, XOR and other equational theories.

%Long term goal: the ability to just state what terms should be protected without having to specify the
%Using custom heuristics, we prove secrecy of all keys involved in the protocol and authentication properties for participant and group keys in platoons up to three vehicles. 

% 0.5 page

\bibliographystyle{IEEEtran}
\bibliography{biblio}

\appendices
\section{Authenticity relation}
In this appendix we prove that the authenticity relation defined in Section~\ref{sec:key_ordering} is transitive. We recall the following definitions:

\begin{itemize}
    \item \textit{Compromising} a term $p$ means either revealing it or being able to generate a $p'$ that will be accepted by other nodes as $p$. We assume that both of these actions (revealing and generating a new key) can be performed infinitely often.
    \item If $\mathcal{K}$ is a set of keys, then the 
    authenticity dependency relation $\dashrightarrow\ \subseteq \mathcal{K} \times \mathcal{K}$ is is defined so that for two keys $k_A, k_B \in \mathcal{K}$, $k_A \dashrightarrow k_B$ holds if compromising $k_B$ allows the attacker to create another key $k'_A$ that will be accepted by the other nodes as the legitimate $k_A$.
\end{itemize}
We now proceed to prove that $\dashrightarrow\subseteq \mathcal{K} \times \mathcal{K}$ is transitive.
\begin{proof}
    Assume that there exists keys $k_A, k_B, k_C \in \mathcal{K}$ such that $k_A \dashrightarrow k_B$ and $k_B \dashrightarrow k_C$. To prove transitivity, we then must show that $k_A \dashrightarrow k_C$. Assume that $k_C$ has been compromised, then by the second definition above, the attacker can create another key $k_B'$ that will be accepted by other nodes as $k_B$. By the first definition, this means that $k_B$ is compromised. Since $k_B$ is compromised and $k_A \dashrightarrow k_B$, then the attacker can create a key $k_A'$ that will be accepted by other nodes as $k_A$. Thus $k_A \dashrightarrow k_C$.
\end{proof}

\section{5G Handover Graphs}
\label{app:handover}

\begin{figure}
    \centering
    \includegraphics[width=\columnwidth]{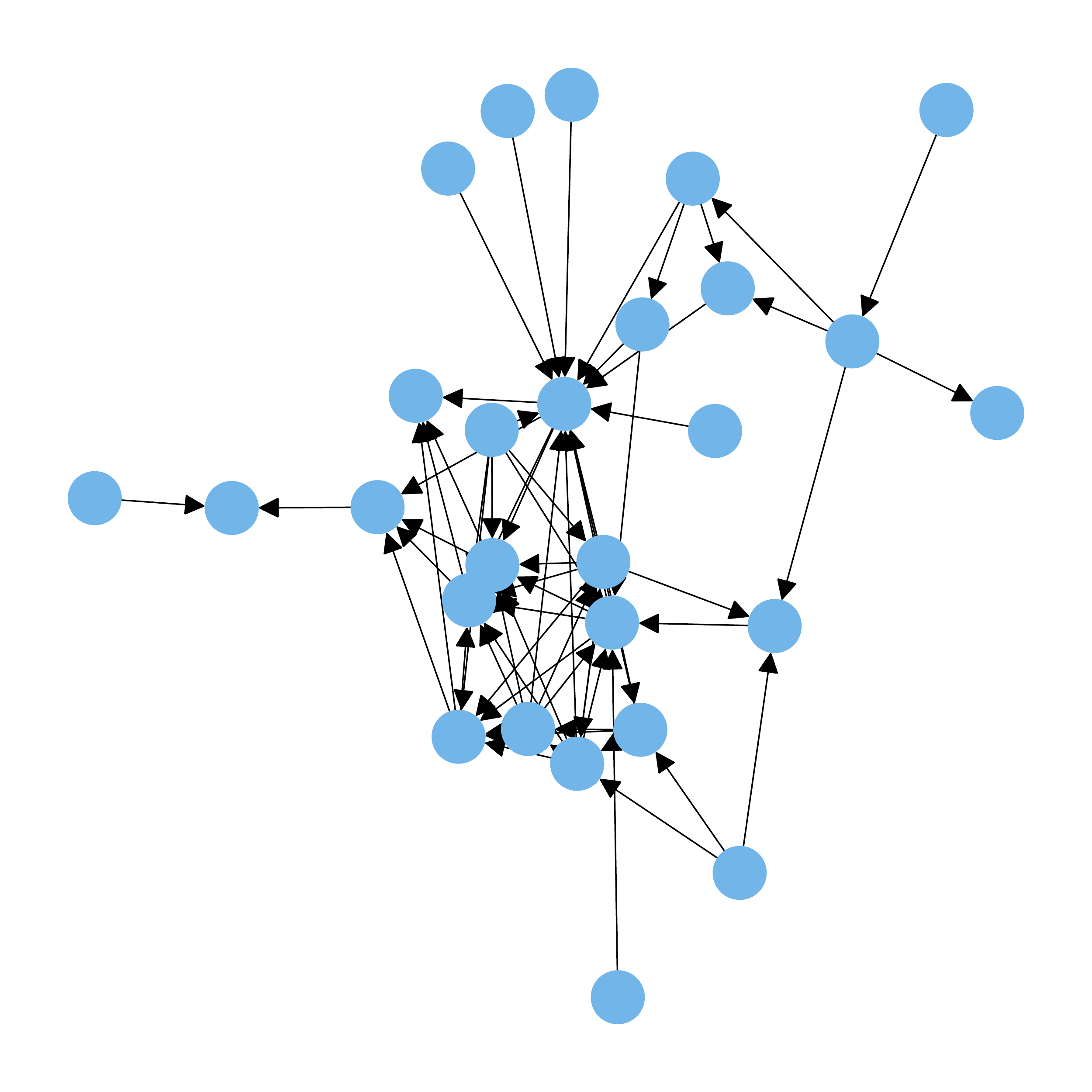}
    \caption{Approximation of the dependency graph from the N2-based inter-RAN variant of 5G handover.}
    \label{fig:app_handover_all}
\end{figure}

In this appendix we present approximations of the key dependencies extracted automatically with our tool from a 5G handover model~\cite{10.1145/3448300.3467823}. In addition to the steps described in Section~\ref{sec:dep_extractor}, a custom key derivation function (KDF with arity 2) present in the 5G handover models must be considered. Given a term $k = \text{KDF}(a, b)$, then it holds that both $a$ and $b$ must be known by an attacker in order to construct $k$. This type of conjunctive dependency is not supported by our current dependency relation, hence we approximate it in a pessimistic (but safe) manner by creating two separate secrecy dependencies $k \rightarrow a$ and $k \rightarrow b$. Intuitively, we state that an attacker could construct $k$ by learning either $a$ or $b$, whereas in reality it must know both terms. This approximation (in addition to our secrecy and authentication relations presented in this work) of the (N2-based inter-RAN) 5G handover model resulted in the graph illustrated in Fig.~\ref{fig:app_handover_all}, which is unlabelled for simplifying the presentation.

\begin{figure}
    \centering
    \includegraphics[width=0.9\columnwidth]{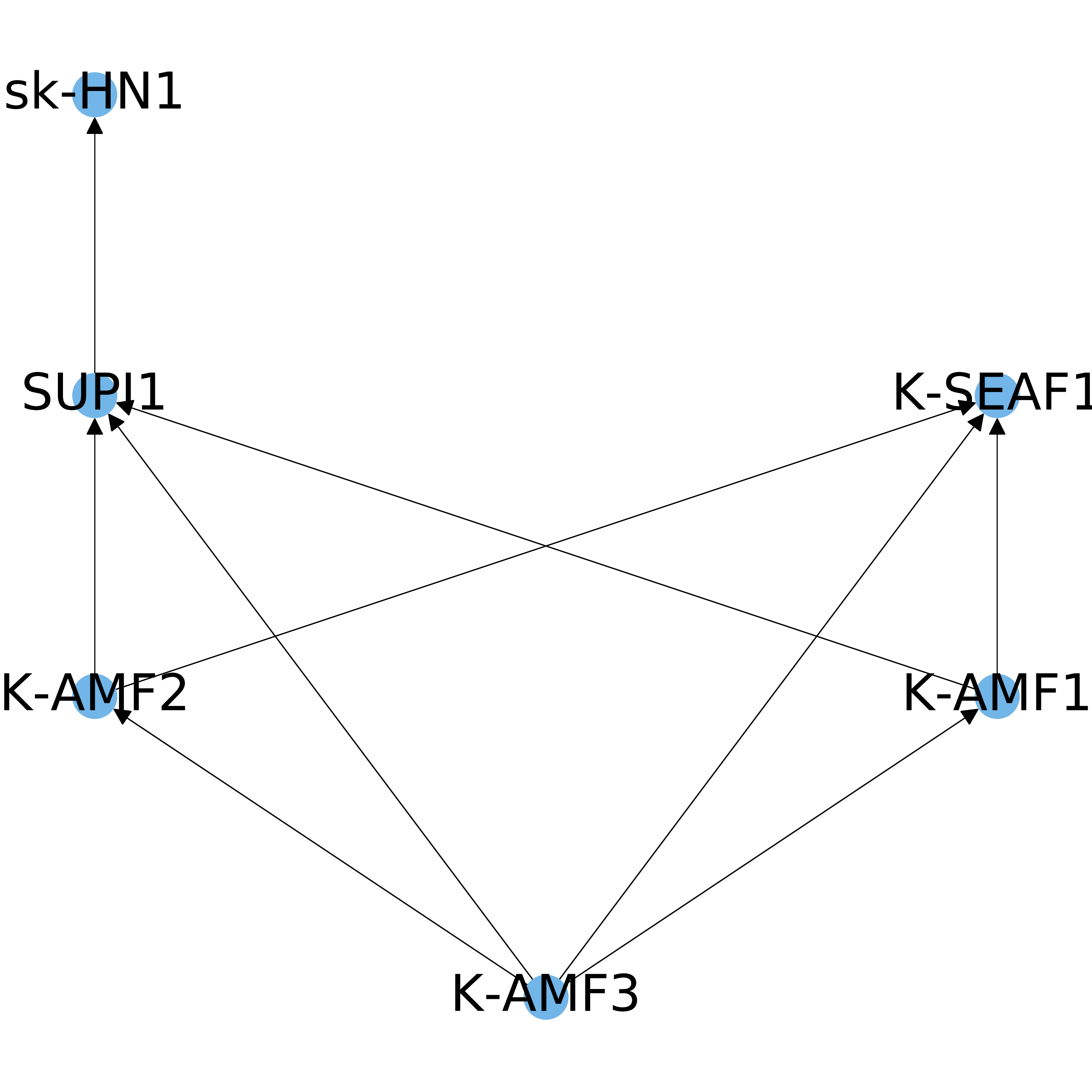}
    \caption{Subgraph of dependencies extracted for target secret 'K-AMF3'.}
    \label{fig:app_handover_1}
\end{figure}

In order to give a concrete example of the structure of this relation, we present in Fig.~\ref{fig:app_handover_1} a subgraph of Fig.~\ref{fig:app_handover_all} which includes the all dependencies originating from a chosen term 'K-AMF3'. 
%To make a distinction between distinct classes of keys, we append an identifier to their labels. 
The secrecy of derived 'K-AMF' keys is one of the verified properties of the 5G handover analysis. A 'K-AMF' can either be derived directly from keys 'SUPI' and 'K-SEAF', or from another 'K-AMF' itself (we refer the reader to the paper~\cite{10.1145/3448300.3467823} for more details). Fig.~\ref{fig:app_handover_1} shows classes of keys (there are several distinct classes of 'K-AMF') and their relations. In addition to the dependencies described earlier, the graph also includes $\text{SUPI} \rightarrow \text{sk-HN}$ due to the asymmetric encryption of 'SUPI' with the public key of the home network $pk(\text{sk-HN})$.

\end{document}